\documentclass[a4paper,fleqn,usenatbib]{mnras}
\usepackage[T1]{fontenc}
\usepackage{ae,aecompl}

\usepackage{graphicx}	
\usepackage{amsmath}	
\usepackage{amssymb}	
\usepackage{xcolor}
\usepackage{tikz}

\usepackage{xcolor} 
\definecolor{darkgreen}{RGB}{0,100,0}

\newcommand{\eht}{\overline}    
\newcommand{\fht}{\widetilde}

\def\av#1{\overline{#1}}
\def\fav#1{\widetilde{#1}}

\def\erho{\eht{\rho}}

\newcommand{\elh}[1]{^{#1}\mbox{H}}
\newcommand{\elhe}[1]{^{#1}\mbox{He}}
\newcommand{\elc}[1]{^{#1}\mbox{C}}
\newcommand{\elo}[1]{^{#1}\mbox{O}}
\newcommand{\elne}[1]{^{#1}\mbox{Ne}}
\newcommand{\elna}[1]{^{#1}\mbox{Na}}
\newcommand{\elmg}[1]{^{#1}\mbox{Mg}}
\newcommand{\elsi}[1]{^{#1}\mbox{Si}}
\newcommand{\elp}[1]{^{#1}\mbox{P}}
\newcommand{\els}[1]{^{#1}\mbox{S}}
\newcommand{\elcl}[1]{^{#1}\mbox{Cl}}
\newcommand{\elar}[1]{^{#1}\mbox{Ar}}

\title[Fine Structure of the Convective-Reactive Zone in a Supernova Progenitor Star]{Structure of Convective-Reactive Zone in a Supernova Progenitor}

\author[M. Moc\'ak et al.]{M. Moc\'ak$^{1}$\thanks{E-mail: miroslav.mocak@gmail.com}, C. Meakin$^{1}$, F. Rizzuti$^{2}$, S. W. Campbell$^{3}$, V. Varma$^{4}$, R. Hirschi$^{4,5}$, C. Georgy$^{6}$,
\newauthor and W.D. Arnett$^{7}$
\\
$^{1}$Aster Scientific, Pasadena, CA 91104 USA\\
$^{2}$Heidelberger Institut f{\"u}r Theoretische Studien, Schloss-Wolfsbrunnenweg 35, D-69118 Heidelberg, Germany\\
$^{3}$School of Physics and Astronomy, Monash University, Clayton, Victoria, Australia\\
$^{4}$Astrophysics Group, Lennard-Jones Laboratories, Keele University, Keele ST5 5BG, UK\\
$^{5}$Kavli IPMU (WPI), University of Tokyo, 5-1-5 Kashiwanoha, Kashiwa 277-8583, Japan\\
$^{6}$Department of Astronomy, University of Geneva, Chemin Pegasi 51, CH-1290 Versoix, Switzerland \\
$^{7}$Steward Observatory, University of Arizona, 933 N. Cherry Avenue, Tucson AZ 85721, USA}

\date{in preparation for MNRAS}

\pubyear{2025}

\begin{document}
\label{firstpage}
\pagerange{\pageref{firstpage}--\pageref{lastpage}}
\maketitle

\begin{abstract}
Convective-reactive events are phases of stellar evolution where turbulent mixing and nuclear burning directly compete, since their timescales become similar within convection zones. During these events the convection zone structure is shaped by a complex and dynamic interaction of convective transport and nuclear burning of individual chemical elements, with plasma streams connecting the entrainment regions with the burning layers. Here we analyse a 3D hydrodynamic simulation of an oxygen--neon shell merger in a massive pre-supernova star. We study the emergent structure using the Reynolds-Averaged Navier-Stokes (RANS) mean-field composition transport equation. We find that the merged convective zone develops a complex, turbulence-driven mixing structure of multiple nested convective-reactive shells, all contained within the single convection zone. Key controlling factors include not only the mean stratification and mixing, but also emergent collective behaviour of involved nuclear reactions for every chemical isotope. Almost all layers show a quasi-steady balance between burning and mixing. We categorise the various layers, and discuss/contrast with 1D stellar evolution code treatments, highlighting some implications.
\end{abstract}


\def\mycmd{1}


\section{Introduction}
\label{s-intro}

The cores of massive pre-supernova stars host a complex hierarchy of burning shells interlaced with convective and radiative zones \citep{Hirschi2004,Yoshida2019}. Recent three-dimensional (3D) hydrodynamics simulations reveal that these shells are prone to mergers due to turbulent entrainment at convective boundaries, particularly in late-stage massive stars where burning layers become weakly stratified \citep{Andrassy2020, Yadav2020}. For instance, studies of neon-oxygen shell interactions demonstrate how entrainment erodes stable layers, enabling the formation of merged convective regions with distinct nucleosynthetic signatures. These events create conditions where nuclear burning timescales become comparable to turbulent transport timescales, governed by local temperature, density and composition at entrainment interfaces and across convection zones. Advanced analysis further demonstrates how entrainment introduces spatially heterogeneous burning regimes within single convection zones \citep{Mocak2018}. Similar multi-scale burning structures emerge in silicon-oxygen shells and carbon-burning regions \citep{Rizzuti2024}, suggesting convective-reactive coupling is ubiquitous in late-stage stellar interiors. According to \citet{Collins2018}, 40$\%$ of supernova progenitors between $16-26 \ \mathrm{M_\odot}$ experience oxygen-neon burning in a convective shell as a result of shell merger shortly before collapse.

Recent observations of supernova remnants indicate, that such shell mergers influence nucleosynthetic yields for intermediate chemical elements (as Ne, Mg and Si) in about 10$\%$ of core collapse supernovae \citep{Matsunaga2024,Sato2024}. These findings underscore the necessity of integrating 3D hydrodynamic constraints into stellar evolution models, particularly for predicting nucleosynthesis products. As computational capabilities advance, combining RANS-based or Reynolds-Averaged Navier-Stokes turbulence analysis \citep{Mocak2014arxiv} with multi-physics simulations of burning shells will be critical for resolving the interplay between nuclear physics and turbulent transport in stellar convection zones. RANS-based turbulence analysis \citep{Pope2000,Chassaing2010,Besnard1992} has been applied to stellar convection and reacting multi-fluid systems in several other works too see e.g. \citet{VialletMeakin2013,amvclm15,ArnettMeakin2009,ArnettMeakin2016}.

\begin{figure*}
\includegraphics[width=0.49\hsize]{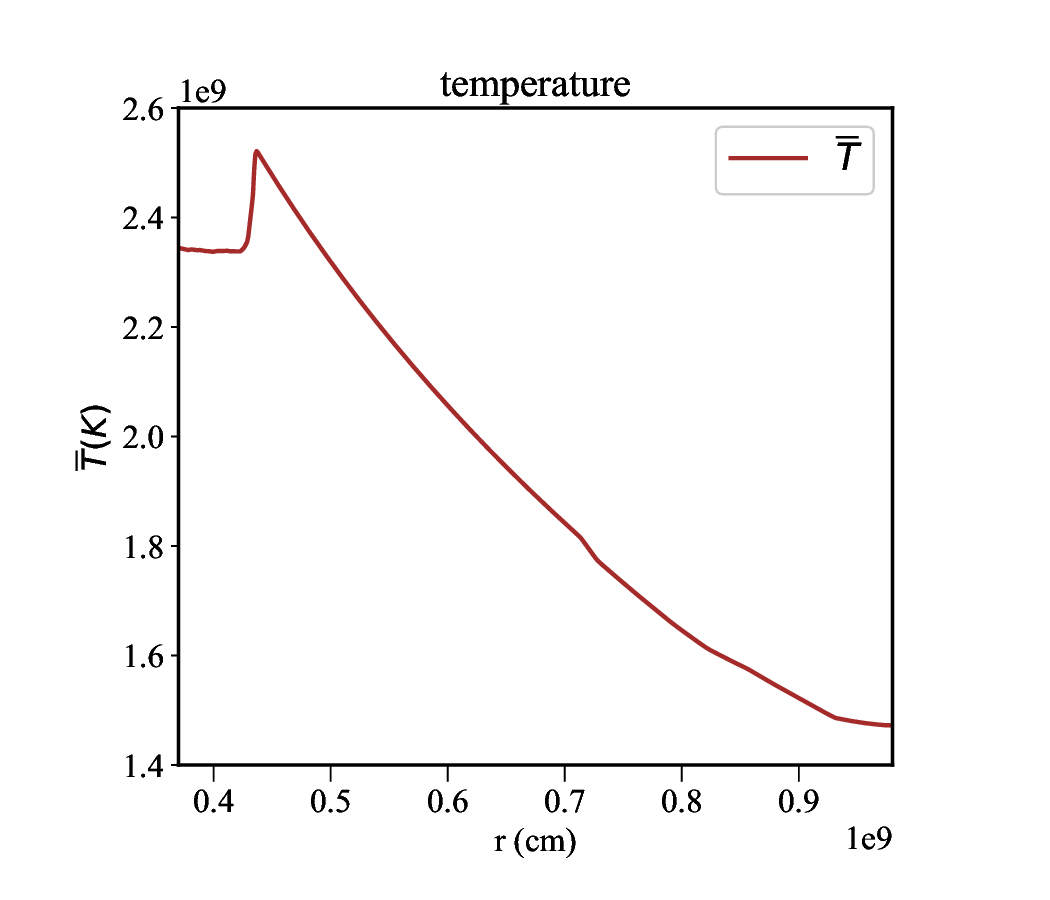}
\includegraphics[width=0.49\hsize]{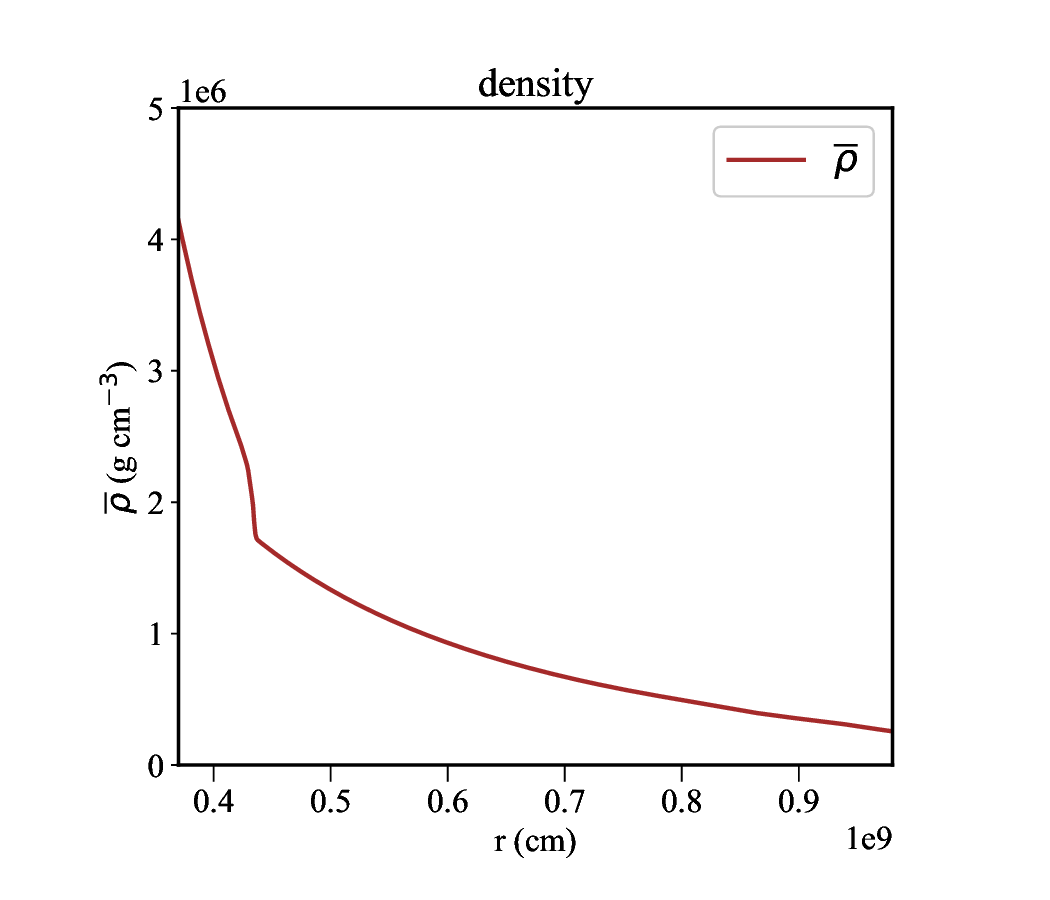}
    \caption{Initial model profiles from the 1D TYCHO progenitor used for the 3D mapping. (Left): Temperature, (Right): Density}
\label{fig:oburn_ini}
\end{figure*}

The focus of the current study is primarily on the analysis of RANS composition transport equations\footnote{It can also be seen as dynamic mean-field continuity equation.} for individual chemical isotopes in our 3D nuclear-hydrodynamics simulation of a burning zone in a massive star. The paper is organized as follows: In Section~\ref{sect:simulation-model} we describe the 3D simulation investigated in this paper, including its initial conditions and the nuclear network utilized. In Section~\ref{sect:rans-analysis} we describe our analysis method for the 3D data, which is based on the RANS approach, including how we obtain the required mean fields. In Section~\ref{sect:results} we present the results of this analysis. We conclude with a discussion and conclusions in Section~\ref{sect:discussion} and \ref{sect:conclusion}. We note that we provide open access to our codes and supporting data products via public repositories and online dashboards upon request.

\section{Initial Model, 3D Hydrodynamic Simulation and Nuclear Networks}
\label{sect:simulation-model}

\begin{table*}
\caption{3D Simulation properties. Columns are: Grid resolution, extension of grid in radial and angular dimensions ($\Delta r$, $\Delta \theta, \Delta \Phi$), utilized number of isotopes in nuclear network (number of species), size of convection zone L$_\mathrm{cvz} = ln (P^{b}_\mathrm{conv}/P^{t}_\mathrm{conv})$ in pressure scale heights H$_p = -P / \partial_r P $, where $P^{b}_\mathrm{conv}$ and $P^{t}_\mathrm{conv}$ is mean pressure at bottom and top of convection zone, averaging timescale of mean fields analysis $\Delta t_\mathrm{av}$; global rms velocity $v_\mathrm{rms} = \sqrt{2 K_{tot}/M}$ ($K_{tot}$ is total turbulent kinetic energy in the convection zone and $M$ mass contained in the convection zone) and convective turnover timescale $\tau_\mathrm{conv} = 2 \ l_{cvz}/v_\mathrm{rms}$, where $l_{cvz}$ is size of the convection zone in $cm$. Mean Mach number in the convection zone is Mach $= \sqrt{u_r^2 / c_s^2}$ where $c_s^2 = \Gamma_1 P/\rho$, effective Reynolds number is $Re = v_{rms} l_{cvz}/ \nu_{eff} \sim N_r^{4/3} $, where $\nu_{eff} \approx \Delta r \Delta u$ is effective viscosity and $N_r$ is number of grid points across the convection zone. All values were obtained at the central simulation time (2221 s) around which we perform all subsequent time-averaging. Note: $Re$ is an effective/estimated value for ILES and should be regarded as order-of-magnitude rather than precise.}
\label{tab:3d-models}
\begin{tabular}{lcccccccccccc}
\hline
Grid & $\Delta r$ & $\Delta \theta$ & $\Delta \phi$ & No. species
& L$_\mathrm{conv}$ (H$_p$) & $\Delta t_\mathrm{av}$ (s) & $v_\mathrm{rms}$  ($10^7$ cm/s)
& $\tau_\mathrm{conv}$  (s) &  $\mbox{Mach}$ & $\mbox{Re}$\\
\hline
512x128x128 & $7\times 10^8$ cm & $45^o$ & $45^o$ & 14 & $2.3$ & $1400$ & 1.1 & $97.4$ & 0.016 & 2438 \\
\hline
\end{tabular}
\end{table*}

Our initial 3D model is an oxygen-burning structure from a pre-supernova massive stellar core computed by the 1D stellar evolution code TYCHO \citep{YoungArnett2005,ArnettMeakin2010}. The 1D progenitor is a $23 \ \mathrm{M_\odot}$ star from which we excise the O-burning shell and adjacent layers; this provides the background stratification (T, $\rho$, composition) used for the mapping to 3D. We refer to \citet{Mocak2018} for additional details of this initial structure and nuclear energetics. 

The model (Fig.~\ref{fig:oburn_ini}) is mapped onto a grid of the hydrodynamical code PROMPI \citep{MeakinArnett2007}, which we use to compute our 3D models. PROMPI has been used in a number of prior stellar convection and shell-burning studies (e.g. \citealt{MeakinArnett2006,MeakinArnett2007,Mocak2018}). The code is capable of performing multidimensional compressible simulations using the finite-volume PPM method with an arbitrary number of chemical isotopes. The characteristics of our 3D simulation are summarized in Table~\ref{tab:3d-models}. This setup follows \citet{Mocak2018} in its underlying 1D progenitor and radial extent, but differs by using a higher resolution grid (512$\times$128$\times$128), a reduced 14-isotope network, a wider angular wedge (45$^{\circ}\times$45$^{\circ}$), and a longer simulated time ($\gtrsim 2000$~s) to reach a quasi-steady convective-reactive state. Using a smaller nuclear network reduces computational cost, enabling the higher grid resolutions and longer simulations presented in this study.

The network contained the following 14 isotopes: $neutrons$, $\elh{1}$, $\elhe{4}$, $\elc{12}$, $\elo{16}$, $\elne{20}$, $\elna{23}$, $\elmg{24}$, $\elsi{28}$, $\elp{31}$, $\els{32}$, $\els{34}$, $\elcl{35}$, $\elar{36}$. In a comparison to our more complete 25 isotope network  simulation \citep{Mocak2018} we found that our 14-isotope simulation lead to similar conclusions about transport and composition/burning, for most overlapping species.\footnote{This direct comparison was performed between 3D simulations with the same radial and angular extent but lower resolution of 128$\times$64$\times$64.}
We caution that reducing the network alters some of the details of the nucleosynthesis; however in this work we focus on the transport--burning balance and layer topology rather than the finer points of the nucleosynthesis. Whereas the initial mass fractions of all chemical isotopes in our 3D model match those of the initial 1D structure up to $\elcl{35}$, this is not the case for $\elar{36}$ which is used as a closure isotope. Its value is derived using the following formula $\elar{36}$ = 1 - $\sum_{\elh{1}}^{\elcl{35}} X_i$. Hence, further conclusions about $\elar{36}$ from our simulation will not be reliable and this isotope is not discussed further. The network includes $\text{78}$ reactions (React.~\ref{react:cl35_1}~$-$ React.~\ref{react:neut_1}) plus four for the closure species $\elar{36}$ i.e. $\elhe{4} + \els{32} \longleftrightarrow \ \elar{36}$ and $\elh{1} + \elcl{35} \longleftrightarrow \ \elar{36}$. All nuclear reaction rates were obtained from the {\it reaclib} library by Friedel Thielemann \citep{raucherthielemann2000}.

\begin{figure*}
\includegraphics[width=0.49\hsize]{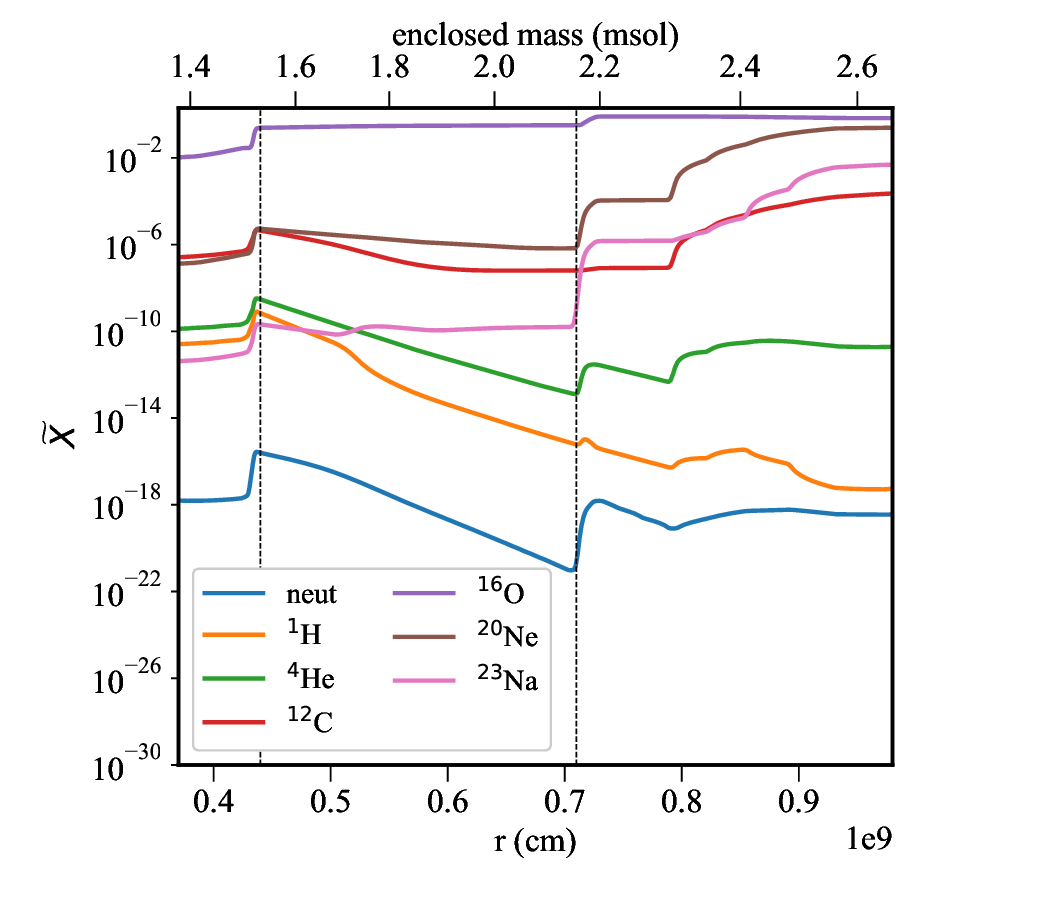}
\includegraphics[width=0.49\hsize]{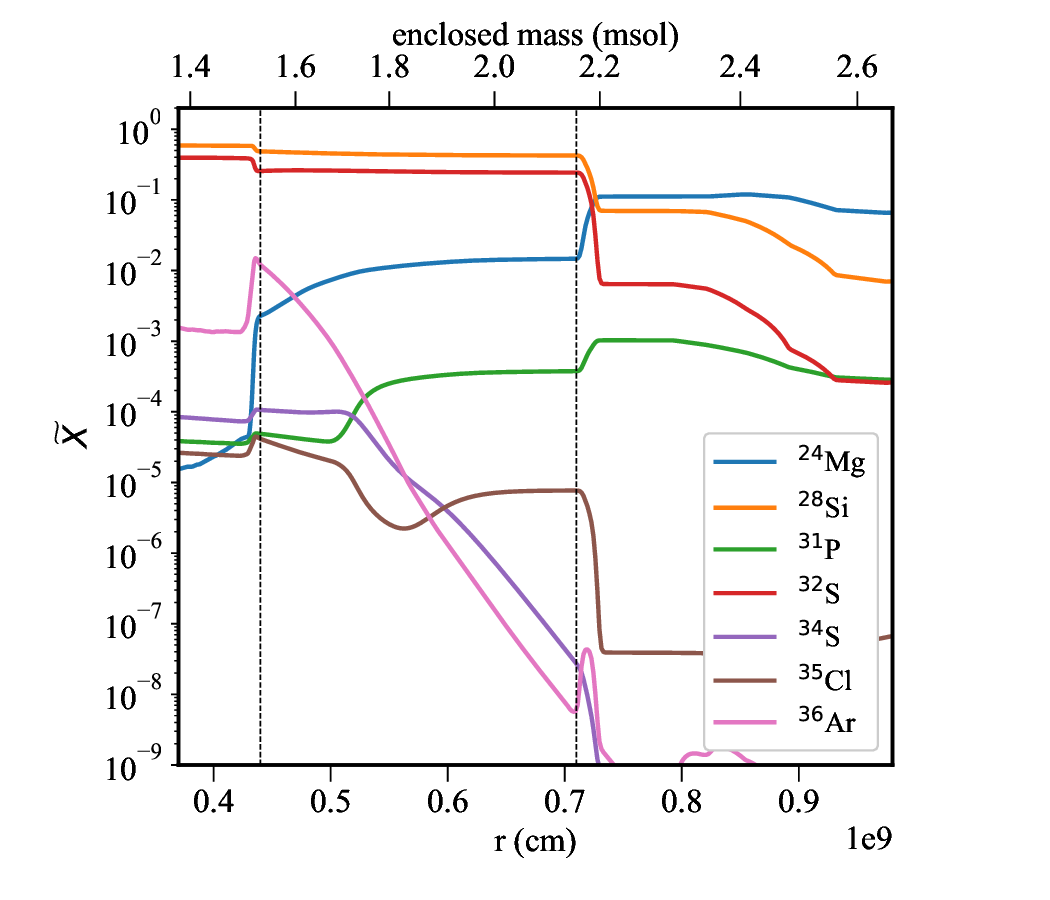} \\

\includegraphics[width=0.49\hsize]{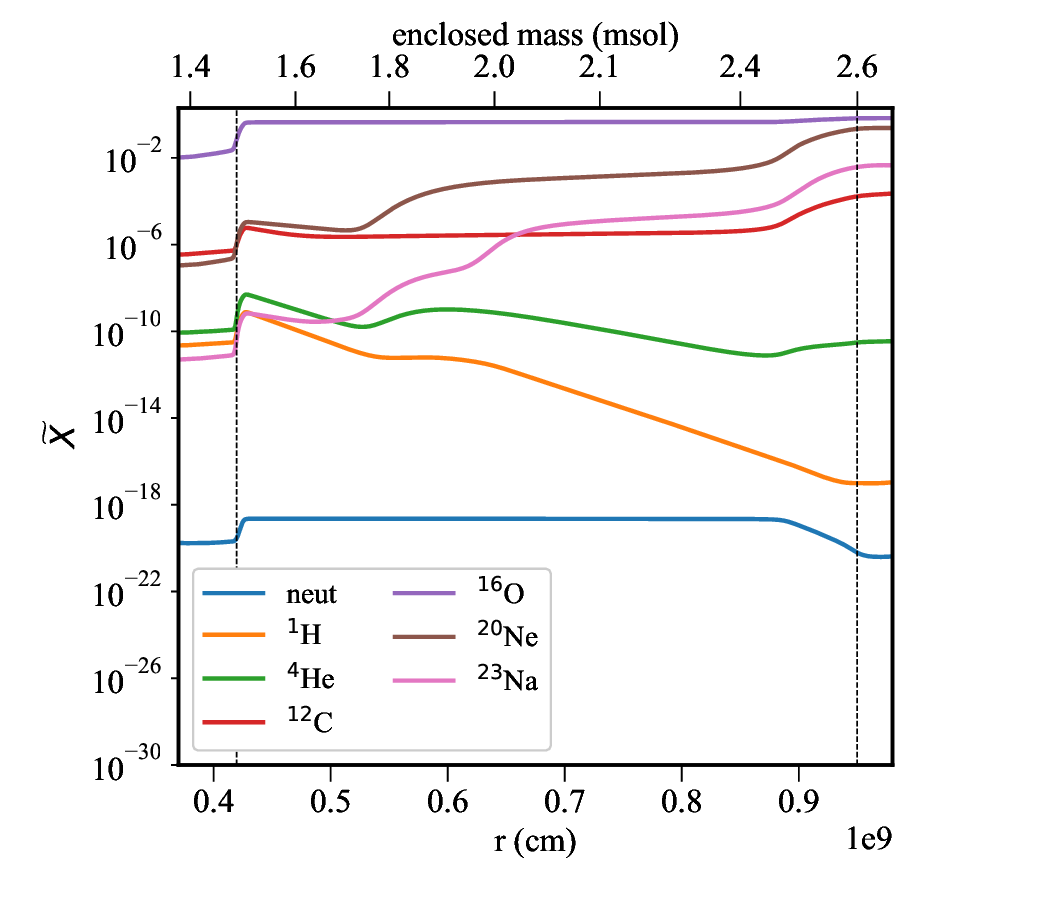}
\includegraphics[width=0.49\hsize]{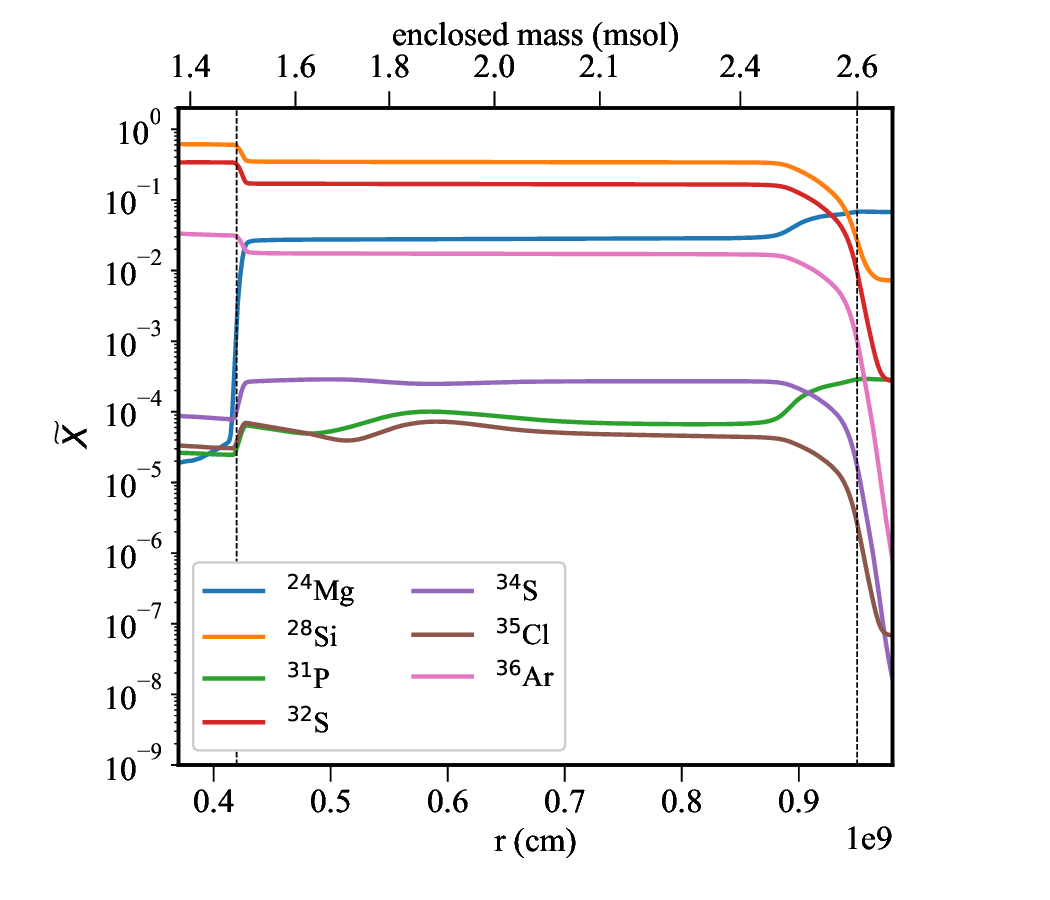}
\caption{Initial composition profiles (upper panels) and composition profiles after 23 turnovers at around 2221 seconds (lower panels). Vertical lines mark convection boundaries defined in Sec.~\ref{sect:timescales} as $r_b^t, r_b^c$. Profiles are snapshots (not time-averaged over convective turnover timescales. We note that these horizontally averaged abundance profiles can appear comparatively flat in 3D for many species; the ``layering'' discussed in this work is diagnosed primarily from the transport balance in Eq.~\ref{eq:transport-comp2} (i.e. from $\partial_t\fav{X}_i$, $\fav{\dot{X}}_i^{\rm nuc}$ and $-(1/\erho)\nabla_r f_i$), rather than from $\fav{X}_i$ alone.}
\label{fig:X_evolved}
\end{figure*}

\section{Analysis Method}
\label{sect:rans-analysis}

Our results are based on RANS analysis of the mean-field transport equation for each of the chemical isotopes we included in our 3D simulation. Such equations are ideal for the analysis of convective-reactive layers, where the transported and burned amount of isotopes is similar.

\subsection{Transport Equation}
\label{sec:transport_equation}

Applying RANS averaging rules \citep{Mocak2014arxiv} on the instantaneous continuity equation for mass fraction of isotope $i$ in spherical geometry,

\begin{align}
\partial_{t} \big(\rho X_{i}\big) =  -\nabla \cdot ( \rho {\bf u} X_i ) + \rho \dot{X}_{i}^{\rm nuc}
\label{eq:cont1}
\end{align}

and having additional implicit numerical residual $res$ in mind (explained in next paragraphs), we obtain the following 1D transport equation

\begin{align}
\partial_t \fav{X}_i  = & +\fav{\dot{X}}_i^{\rm nuc} - (1/\erho)\nabla_r f_i -\fav{u}_r \partial_r \fav{X}_i + res
\label{eq:transport-comp2}
\end{align}

\noindent where $X_i$ is mass fraction of chemical isotope $i$, $\rho$ is density, ${\bf u} = {\bf u} (u_r,u_\theta, u_\phi)$ is the velocity vector, $\nabla$ is the divergence operator, $\dot{X}_i^{\rm nuc}$ is the rate of nuclear burning of $i$, and $f_i = \eht{\rho}\fht{X''_i u''_r}$ is turbulent flux of isotope $i$. $\nabla_r (.) = (1/r^2)\partial_r r^2 (.)$ is the radial divergence operator. The averaging procedure is explained in Appendix \ref{appendix:averaging}.

\begin{figure}
\centering
\resizebox{\columnwidth}{!}{
\begin{tikzpicture}[x=1pt,y=1pt,>=latex]
  \path[use as bounding box] (0,0) rectangle (430,240);

  \node[align=center,font=\bfseries] (type) at (160,125) {\fontsize{28}{30}\selectfont Type};
  \node[align=center,font=\bfseries] (ii)   at (215,125) {\fontsize{28}{30}\selectfont II};
  \node[align=center,font=\bfseries] (star) at (234,132) {\fontsize{28}{28}\selectfont $^{\boldsymbol{*}}$};
  \node[align=center,font=\bfseries] (pm)   at (265,125) {\fontsize{28}{30}\selectfont $\boldsymbol{+/-}$};

  \tikzset{
    lab/.style={font=\small},
    arr/.style={line width=1.4pt, -latex, shorten <=8pt, shorten >=10pt},
  }

  \node[lab,font=\Huge] (layers) at (75,210) {Number Of Layers};
  \node[lab,font=\Huge] (entr)   at (355,210) {Significant Entrainment};

  \node[lab,font=\Huge] (chg) at (265,35) {Increase/Decrease of $X$};

  \draw[arr] (layers.south) -- ++(0,-18) -| (ii.north);
  \draw[arr] (entr.south)   -- ++(0,-18) -| (star.north);
  \draw[arr] (chg.north)    -- (pm.south);
\end{tikzpicture}}
\caption{Naming convention explainer of the discovered substructures for individual isotopes. The ``Number Of Layers'' specifies how many distinct layers are present; ``Significant Entrainment'' indicates notable mixing at convection boundaries of the isotope from stable layers in or out of the convection zone where $-(1/\erho)\nabla_r f_i \sim \max(\fav{\dot{X}}_i^{\rm nuc})$; and the "Increase/Decrease" of mass fraction refers to changes in the proportion of mass fraction in the middle of the convection zone compared to the initial state. $+$ indicates increase of $\fav{X}_i$, $-$ indicates decrease of $\fav{X}_i$. If $\fav{X}_i$ increases in some layers and decreases in others, the sign refers to the net change evaluated at the mid-convection-zone reference point.}\label{fig:type_explainer}
\end{figure}
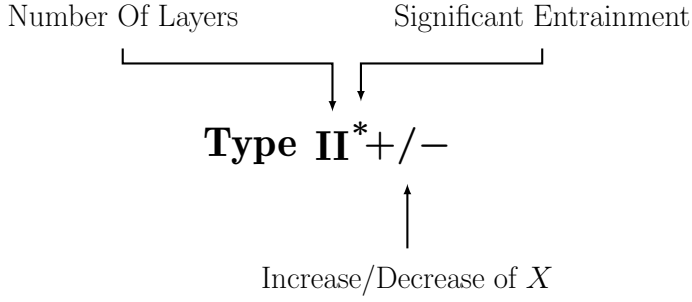

\par The mean-field transport equation (Eq.~\ref{eq:transport-comp2}) tells us that the temporal change of mass fraction of an isotope $i$ in the Eulerian frame of reference,  $\partial_t \fav{X}_i$, is caused by nuclear burning $\fav{\dot{X}}_i^{\rm nuc}$, redistribution due to the turbulent flux $f_i$ by $-(1/\erho) \nabla_r f_i$ and background expansion $\fav{u}_r \partial_r \fav{X}_i$. We define the numerical residual in these equations by $res$, which represents the implicit action of our numeric scheme. This equation is one of the stellar structure equations simulated in 1D \citep{Kippenhahn1990stellar} augmented by changes due to background expansion and turbulent transport. Turbulent transport depends directly on divergence of composition flux, which is an optional term in 1D simulations typically treated as a diffusion (e.g. \citealt{christensen2008astec}). However, its nature in supernova core convection shell mergers is turbulent.

We obtain all the mean fields ($\eht{\rho},\fht{u}_r,\fav{X}_i, f_i, \av{\rho}\fav{\dot{X}}_i^{\rm nuc}$) at runtime of our hydrodynamic code PROMPI and by post-processing with the rans(eXtreme) framework \footnote{\href{https://github.com/mmicromegas/ransX}{https://github.com/mmicromegas/ransX}}, which was programmed in Python in order to streamline RANS analysis of 3D hydrodynamic compressible multi-fluid simulations. 

Depending on values of all these mean-fields, the transport equation results in our simulations showing three main qualitatively different scenarios:

\begin{itemize}
\item Scenario 1: if nuclear burning {\it destruction} ($+\fav{\dot{X}}_i^{\rm nuc}$ is negative) balances the transport $-(1/\erho) \nabla_r f_i$, it means that whatever amount of gas is transported into this layer from beyond will be burned off, so the composition $\fav{X}_i$ remains constant ($\partial_t \fav{X}_i \sim 0$).
\item Scenario 2: This is the opposite of Scenario 1. If the nuclear burning {\it production} ($\fav{\dot{X}}_i^{\rm nuc}$ is positive) balances the transport $-(1/\erho)\nabla_r f_i$, then whatever amount of isotope is produced by the nuclear burning will also be completely replaced by transport from elsewhere, hence the composition mass fraction $\fav{X}_i$ remains constant ($\partial_t \fav{X}_i \sim 0$).
\item Scenario 3: If the time rate of change of the composition density $\partial_t \fav{X}_i$ has a value comparable to either the transport $-(1/\erho)\nabla_r f_i$ or the nuclear burning $+\fav{\dot{X}}_i^{\rm nuc}$, this means that most of the isotope is only being transported around without being affected by nuclear burning, or vice versa. The sign of the value of the term determines whether the material is added or removed. The scenario is clearly evident at convection boundaries, where fresh gas is either removed by entrainment from the high abundance layer, or added from inside the convection zone to the low abundance layers, and $\partial_t \fav{X}_i \sim -(1/\erho)\nabla_r f_i$. 

This species-specific behaviour is not captured by standard 1D prescriptions, where all species are typically mixed together with a single diffusion coefficient. 

\end{itemize}

{\it These scenarios lead to distinct layered substructures for each isotope within our convection zone, identified by regions of either nuclear production or destruction (positive or negative $\fav{\dot{X}}_i^{\rm nuc}$, respectively)}. 

To categorise the various structures observed, we propose a new naming convention for each subtype of the structures according to the key explained in Fig.~\ref{fig:type_explainer}. In words: the label encodes (i) the number of layers (I, II, III, …), (ii) the net change of chemical isotope mass fraction within the merged convective region ($+$ net increase; $-$ net decrease), and (iii) an asterisk if significant boundary entrainment contributes at a level comparable to the peak RANS transport term [$-(1/\erho)\nabla_r f_i$].  Properties of all the structures are summarized at the end of paper in Tab.~\ref{tab:structure_summary}.

We find that the redistribution due to background expansion $\fav{u}_r \partial_r \fav{X}i$ is very weak in all cases because expansion velocities are only around $10^4$ cm/s, so the background evolves only minimally over the relatively short averaging window ($\Delta t\mathrm{av} \sim 1400$ s i.e. secular expansion terms are subdominant compared to turbulent transport, where turbulent velocities are $\sim 10^3$ greater around $10^7$ cm/s.

\subsection{Relevant timescales}
\label{sect:timescales}

Important features of the simulations are strongly dynamic, and it is practical to understand them in terms of some timescales, which we define here.

The convective turnover timescale is
\begin{align}
\tau_\mathrm{conv} = 2 (r^c_\mathrm{t} - r^c_\mathrm{b}) / v_\mathrm{rms},
\end{align}
where $r^c_\mathrm{b}$ and $r^c_\mathrm{t}$ are the radii of the bottom and top convection boundaries, and $v_\mathrm{rms}$ is the rms of the velocity field in the convection zone. The criterion for choosing the radius of the bottom and top of the convective boundary $r^c_\mathrm{b}, r^c_\mathrm{t}$ is taken as where the composition flux drops to practically zero \citet{Canuto1999}. In practice, we define $r^c_\mathrm{b}$ and $r^c_\mathrm{t}$ as the locations where composition flux $|f_i|$ falls below a fixed threshold of $3 \times 10^{-3}$ of its peak magnitude in the convection zone. $v_\mathrm{rms} = \sqrt{2 E_k / M}$ is the rms velocity of the turbulence, where $M$ is its total mass within the convective region and $E_k$ is the flow's total turbulent kinetic energy. The net nuclear burning timescale for isotope $i$ is

\begin{align}
\tau_{\rm nuc}^{i} = \fht{X}_i / \fav{\dot{X}}_i^{\rm nuc}
\label{eq:net-nucl-timescale}
\end{align}

and the (e-folding) turbulent transport timescale of isotope $i$ is

\begin{align}
\tau_{\rm trans}^{i} = \eht{\rho} \fht{X}_i / (\nabla_r f_i).
\label{eq:turbulent-transport-timescale}
\end{align}

The transport timescale and convective turnover timescale are fundamentally different and their values may often differ locally by orders of magnitude.

\section{Results}
\label{sect:results}

Our 3D hydrodynamic simulation of the oxygen burning shell leads, after an initial transient episode, to a merger of the oxygen and adjacent neon burning shell \citep{Mocak2018}. The shell merger significantly affects the compositional structure shown in Fig.~\ref{fig:X_evolved}, which is very different from the initial ``1D'' state. The final compositional structure results from a combination of entrainment from/to stable layers, and newly induced nuclear burning.

\begin{figure}
\includegraphics[width=0.99\hsize]{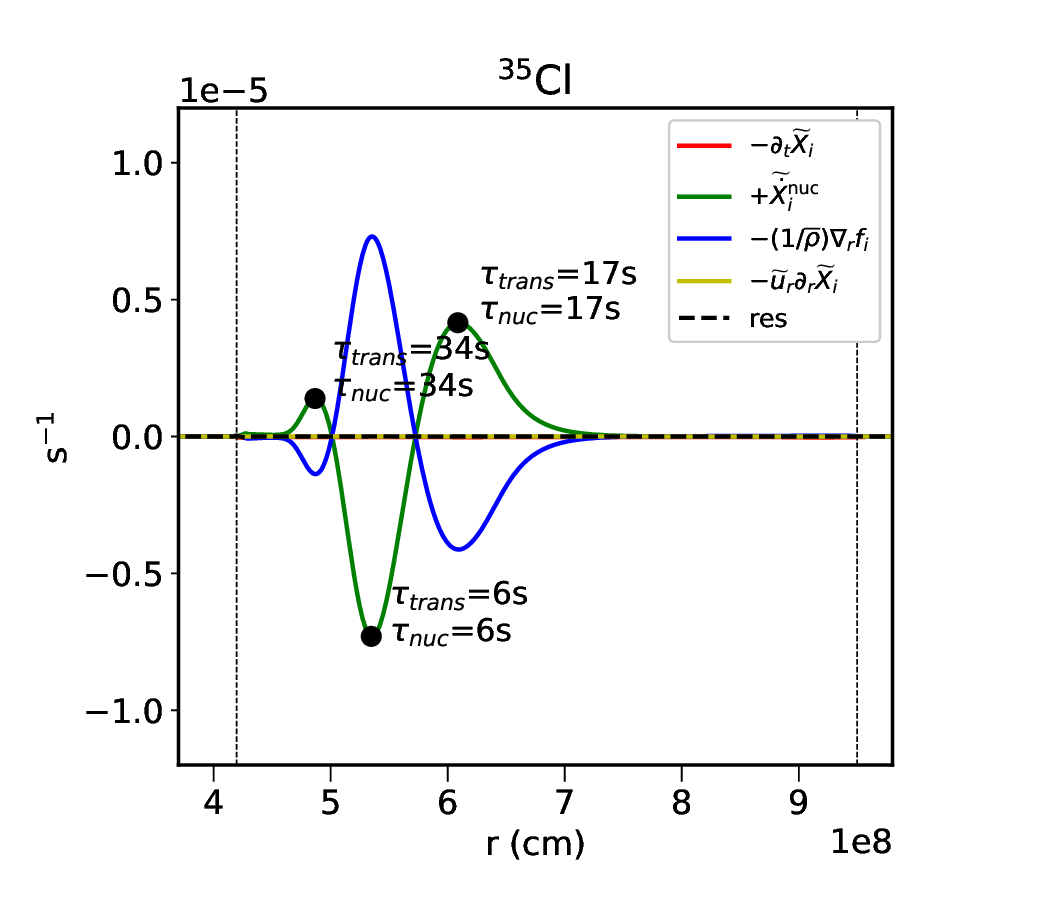}
 \caption{Transport Equation for $\elcl{35}$; spatially and temporally averaged. $\tau_{nuc}$ and $\tau_{trans}$ are nuclear and transport timescales i.e. Eq.\ref{eq:net-nucl-timescale} and Eq.\ref{eq:turbulent-transport-timescale}, respectively.}
\label{fig:ransX_set_ar36_cl35}
\end{figure}

\begin{figure*}
\includegraphics[width=0.49\hsize]{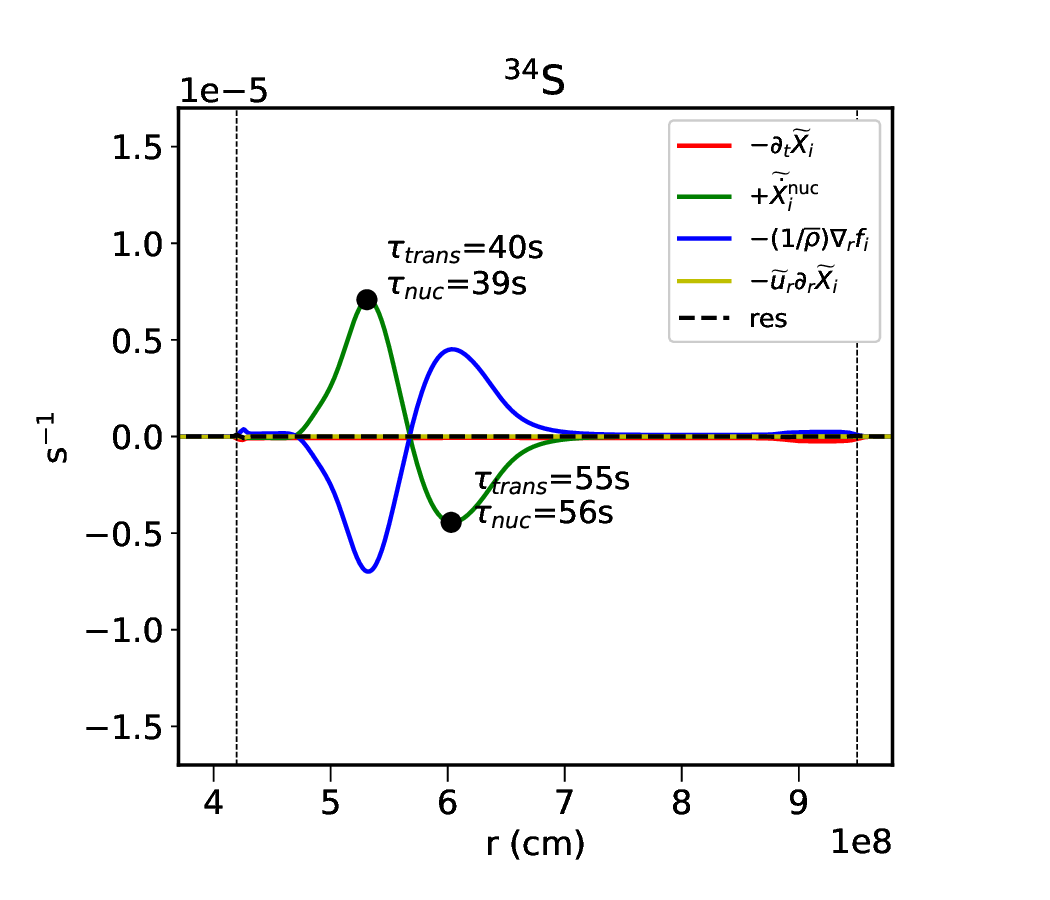}
\includegraphics[width=0.49\hsize]{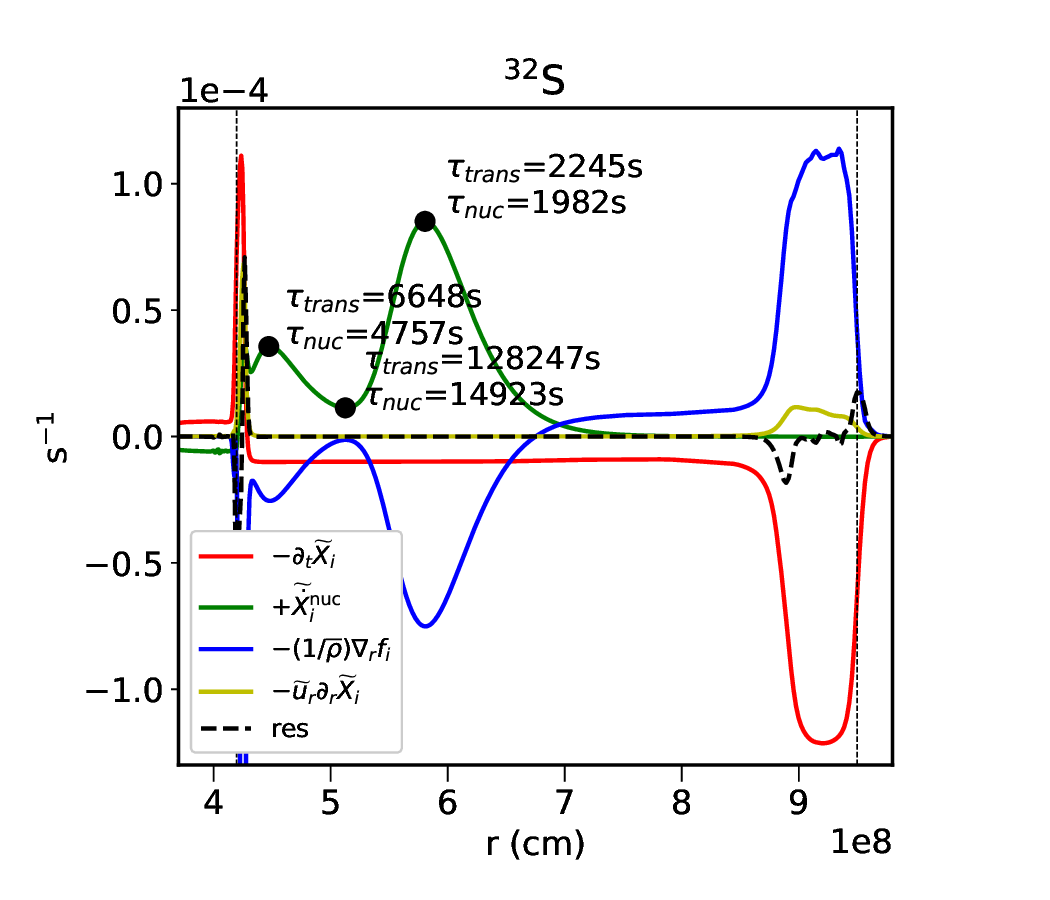}
\caption{Transport Equation for $\els{34}$ (left), $\els{32}$ (right); spatially and temporally averaged. $\tau_{nuc}$ and $\tau_{trans}$ are nuclear and transport timescales i.e. Eq.\ref{eq:net-nucl-timescale} and Eq.\ref{eq:turbulent-transport-timescale}, respectively.}
\label{fig:ransX_set_s34_s32}
\end{figure*}

Based on the dominant physical processes in our convective zone, represented by terms in the transport equation, it it is possible to identify several qualitatively different convective-reactive layers, which are often interleaved (Figures \ref{fig:ransX_set_ar36_cl35}~--~\ref{fig:ransX_set_he4_h1_neut}). These structures are nested within one continuous convection zone (i.e. there are no ``gaps'' in convection); instead, different radial sub-regions exhibit different local balances of transport and burning, for each isotope.

A convective-reactive layer can be defined as a region with a distinct nuclear burning rate of a given isotope balanced by its transport ($+\fav{\dot{X}}_i^{\rm nuc} \sim -(1/\erho)\nabla_r f_i$). This results in similar nuclear and transport timescales, with the Damk\"ohler number ($D_a = \tau_{trans}/ \tau_{nuc}$) approaching unity. We emphasise that the exact location and multiplicity of layers reflect both the dominant nuclear reaction channels and the efficiency of turbulent redistribution; reaction sub-networks can overlap spatially which may affect the number of distinct layers.

\subsection{Fine Structure of the Convection Zone}

The simulated convection zone is mainly composed of $\elo{16}, \elsi{28}, \els{32}, \elmg{24}$ (X > $10^{-2}$), but after merging with the neon shell it is also heavily polluted by $\elne{20}$, $\elc{12}, \elna{23}$ and $\elmg{24}$ from the outer stable layers. The pollution by entrainment from the inner stable layer comes mainly from $\elsi{28}$ and $\els{32}$.

In the following subsections we begin to describe the fine structure of the convection zone by looking at nuclear reactions and the profiles of terms in the transport equations from $\elcl{35}$ down to $\elh{1}$ and neutrons. Where multiple dominant reactions for a given species operate in overlapping radial domains, we note that the layer counts reflect the effective sum of overlapping channels.

\subsection{$\elcl{35}$}
\label{sect:cl35}

Nuclear burning of $\elcl{35}$ (Fig.~\ref{fig:ransX_set_ar36_cl35}) is controlled by Reactions \ref{react:cl35_1}, \ref{react:cl35_2}, \ref{react:cl35_3}, \ref{react:cl35_4}:

\begin{align}
\elhe{4} + \elp{31} \longleftrightarrow & \ \elcl{35} \label{react:cl35_1} \\
 \elh{1} + \els{34} \longleftrightarrow & \ \elcl{35} \label{react:cl35_2} \\
 \elh{1} + \elcl{35} \longleftrightarrow & \ \elar{36} \label{react:cl35_3} \\
 \elh{1} + \elcl{35} \longleftrightarrow & \ \elhe{4} + \els{32} \label{react:cl35_4}
\end{align}

We identify two nuclear production regions ($\fav{\dot{X}}_i^{\rm nuc}$ > 0) with peaks at
radii $4.8 \times 10^8$ cm and $6 \times 10^8$ cm and one destruction region ($\fav{\dot{X}}_i^{\rm nuc}$ < 0) with a peak at radius $5.2 \times 10^8$ cm, with peak nuclear timescales around 34 s, 17 s and 6 s, respectively. The transport timescales are identical for all peaks, which means that $\elcl{35}$ exhibits a three-layer convective-reactive structure, with the two nuclear production regions being the source of $\elcl{35}$ for the single nuclear destruction region in between. The entrainment from stable layers is relatively weak, since $\partial_t (\fav{X}_i)$ is much smaller than $-(1/\erho)\nabla_r f_i$ everywhere. We classify the structure as Type III+ (see Fig.~\ref{fig:type_explainer}; Tab.~\ref{tab:structure_summary}).

\subsection{$\els{34}$}
\label{sect:s34}

Nuclear burning of $\els{34}$ (Fig.\ref{fig:ransX_set_s34_s32}) is controlled by
Reactions \ref{react:s34_1}, \ref{react:s34_2}:

\begin{align}
\elh{1} + \els{34} \longleftrightarrow & \ \elcl{35}\label{react:s34_1} \\
\elh{1} + \els{34} \longleftrightarrow & \ \elhe{4} + \elp{31} \label{react:s34_2}
\end{align}

There is a nuclear production region ($\fav{\dot{X}}_i^{\rm nuc}$ > 0) with peak at radius $5.2 \times 10^8$ cm and a nuclear destruction region ($\fav{\dot{X}}_i^{\rm nuc}$ < 0) with peak at $6 \times 10^8$ cm with nuclear timescales around 39 s and 56 s, respectively. The transport timescales (40 s and 55 s) are almost identical at this location, indicating a double-layered convective-reactive structure, with nuclear production in the first layer being the source of $\els{34}$ to be burned off in the second. Entrainment from stable layers is relatively weak, since $\partial_t (\fav{X}_i)$ is much smaller than $-(1/\erho)\nabla_r f_i$ everywhere. We classify the structure as Type II+ (Tab.~\ref{tab:structure_summary}).

\subsection{$\els{32}$}
\label{sect:s32}

Nuclear burning of $\els{32}$ (Fig.\ref{fig:ransX_set_s34_s32}) is controlled by Reactions \ref{react:s32_1}, \ref{react:s32_2}, \ref{react:s32_3}, \ref{react:s32_4}:

\begin{align}
\elhe{4} + \elsi{28} \longleftrightarrow & \ \els{32} \label{react:s32_1} \\
\elh{1} + \elp{31} \longleftrightarrow & \ \els{32} \label{react:s32_2} \\
 \elhe{4} + \els{32} \longleftrightarrow & \ \elar{36} \label{react:s32_3} \\
\elh{1} + \elcl{35}  \longleftrightarrow & \ \elhe{4} + \els{32} \label{react:s32_4}
\end{align}

\begin{figure*}
\includegraphics[width=0.49\hsize]{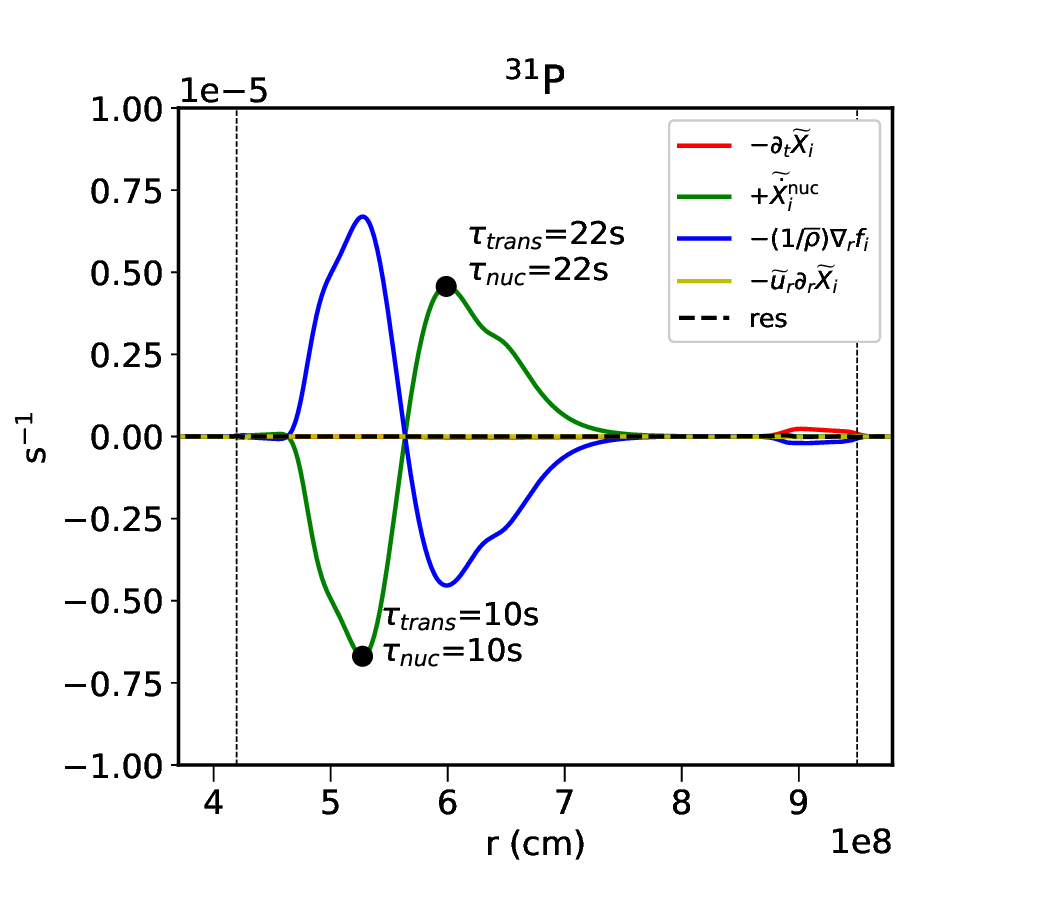}
\includegraphics[width=0.49\hsize]{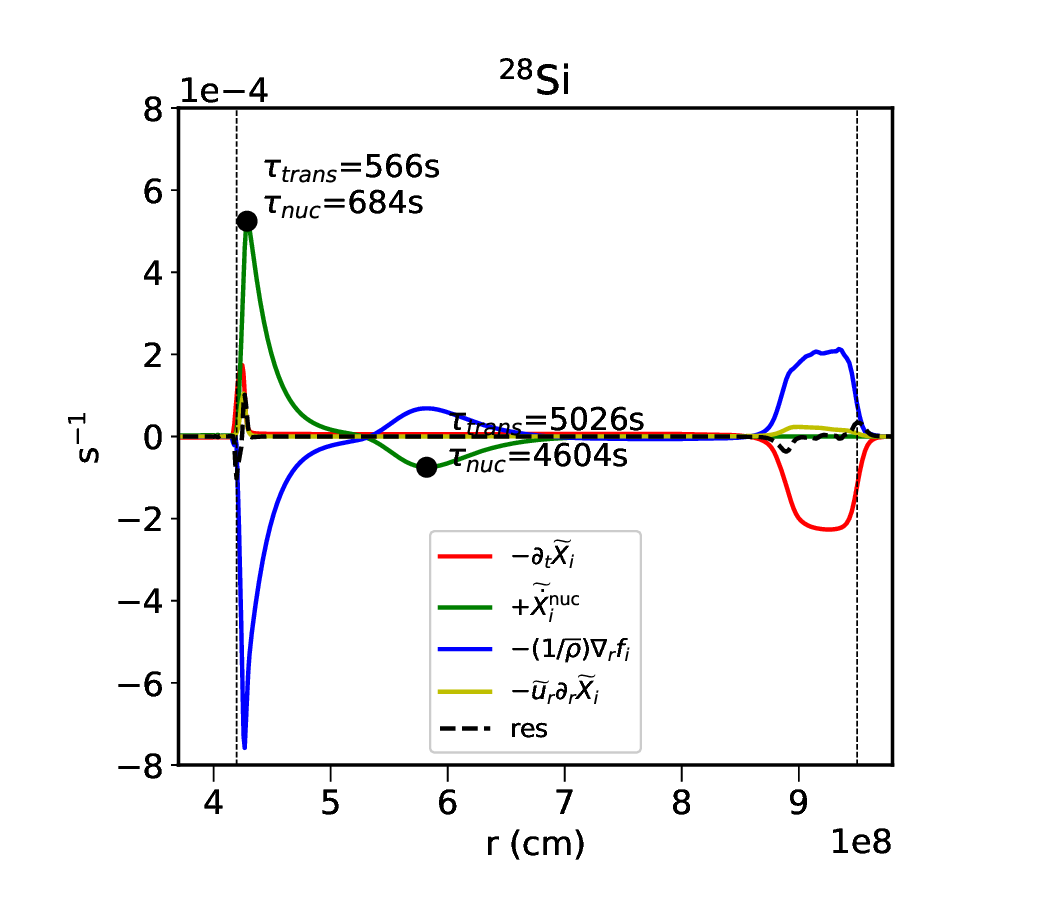}
\caption{Transport Equation for $\elp{31}$ (left) and $\elsi{28}$ (right); spatially and temporally averaged. $\tau_{nuc}$ and $\tau_{trans}$ are nuclear and transport timescales i.e. Eq.\ref{eq:net-nucl-timescale} and Eq.\ref{eq:turbulent-transport-timescale}, respectively.}
\label{fig:ransX_set_p31_si28}
\end{figure*}

We identify a single nuclear production region ($\fav{\dot{X}}_i^{\rm nuc}$ > 0) with two dominant peaks and one valley with decreased nuclear burning around $5 \times 10^8$ cm. 
The nuclear and transport timescales for the first peak at $4.3 \times 10^8$ cm are 4757 s and 6648 s, respectively. The second peak at $6 \times 10^8$ cm has much similar nuclear and transport timescales around 1982 s and 2245 s, respectively, indicating a single-layer convective-reactive shell. Entrainment from the $\els{32}$-rich lower stable layer and the produced $\els{32}$ becomes a major source of contamination for the upper stable layer, where $\partial_t (\fav{X}_i) \sim -(1/\erho)\nabla_r f_i$. We classify the structure as Type I*- (Tab.~\ref{tab:structure_summary}).

\subsection{$\elp{31}$}
\label{sect:p31}

Nuclear burning of $\elp{31}$ (Fig.\ref{fig:ransX_set_p31_si28}) is controlled by
Reactions \ref{react:p31_1}, \ref{react:p31_2}, \ref{react:p31_3}, \ref{react:p31_4},
\ref{react:p31_5}, \ref{react:p31_6}:

\begin{align}
\elh{1} + \elp{31} \longleftrightarrow & \ \els{32} \label{react:p31_1} \\
\elhe{4} + \elp{31}  \longleftrightarrow & \ \elcl{35} \label{react:p31_2}   \\
 \elh{1} + \elp{31} \longleftrightarrow & \ \elhe{4} + \elsi{28} \label{react:p31_3}  \\
 \elh{1} + \els{34} \longleftrightarrow & \ \elhe{4} + \elp{31} \label{react:p31_4} \\
 \elo{16} + \elo{16} \longrightarrow & \ \elh{1} + \elp{31} \label{react:p31_5} \\
 \elc{12} + \elne{20} \longrightarrow & \ \elh{1} + \elp{31} \label{react:p31_6}
\end{align}

We identify one nuclear production region ($\fav{\dot{X}}_i^{\rm nuc}$ > 0) with peak at
radius $6.2 \times 10^8$ cm and nuclear and transport timescales being identical around 22 s. There is also a nuclear destruction region ($\fav{\dot{X}}_i^{\rm nuc}$ < 0) at $5.2 \times 10^8$ cm with nuclear and transport timescales of 10~s. Compared to burning and transport within the convection zone, transport to/from convection boundaries is weak i.e. the term $(1/\erho)\nabla_r f_i$ is much smaller than its counterparts within the zone. These findings indicate a two-layer convective reactive structure. We classify the structure as Type II- (Tab.~\ref{tab:structure_summary}).

\subsection{$\elsi{28}$}
\label{sect:si28}

Nuclear burning of $\elsi{28}$ (Fig.~\ref{fig:ransX_set_p31_si28}) is a combination of photo-disintegration and nuclear fussion shown by Reactions \ref{react:si28_1}, \ref{react:si28_2}, \ref{react:si28_3}, \ref{react:si28_4}, \ref{react:si28_5}:

\begin{align}
\elhe{4} + \elmg{24} \longleftrightarrow & \ \elsi{28} \label{react:si28_1} \\
\elhe{4} + \elsi{28}  \longleftrightarrow & \ \els{32} \label{react:si28_2} \\
 \elh{1} + \elp{31} \longleftrightarrow & \ \elhe{4} + \elsi{28} \label{react:si28_3}  \\
 \elo{16} + \elo{16} \longrightarrow & \ \elhe{4} + \elsi{28} \label{react:si28_4} \\
 \elc{12} + \elne{20} \longrightarrow & \ \elhe{4} + \elsi{28} \label{react:si28_5}
\end{align}

We identify a nuclear production region ($\fav{\dot{X}}_i^{\rm nuc}$ > 0) in a narrow shell around temperature maximum at $4 \times 10^8$ cm and a nuclear destruction region ($\fav{\dot{X}}_i^{\rm nuc}$ < 0) peaking around $5.8 \times 10^8$ cm with nuclear timescales around $684$ s and $4604$ s, respectively. The transport timescales are slightly smaller for the first peak around $566$ s. It means that some of the $\elsi{28}$ is transported faster on shorter timescales before it is processed by nuclear burning because it burns on longer timescales. This effect occurs due to fast entrainment from the underlying $\elsi{28}$-rich stable layer below, leading to its transport through the convection zone to the upper layers. This is indicated by a narrow peak of $\partial_t (\fav{X}_i)$ at $4.2 \times 10^8$ cm. The situation is oposite at the second peak, where the nuclear timescale is smaller than transport timescale by about 8$\%$. These findings indicate a two-layer convective reactive structure. We find a signficant contamination of the upper stable layer by transport from underlaying layers where $\partial_t (\fav{X}_i) \sim -(1/\erho)\nabla_r f_i$ above $9 \times 10^{8}$ cm. We classify the structure as Type II*- (Tab.~\ref{tab:structure_summary}).

\begin{figure*}
\includegraphics[width=0.49\hsize]{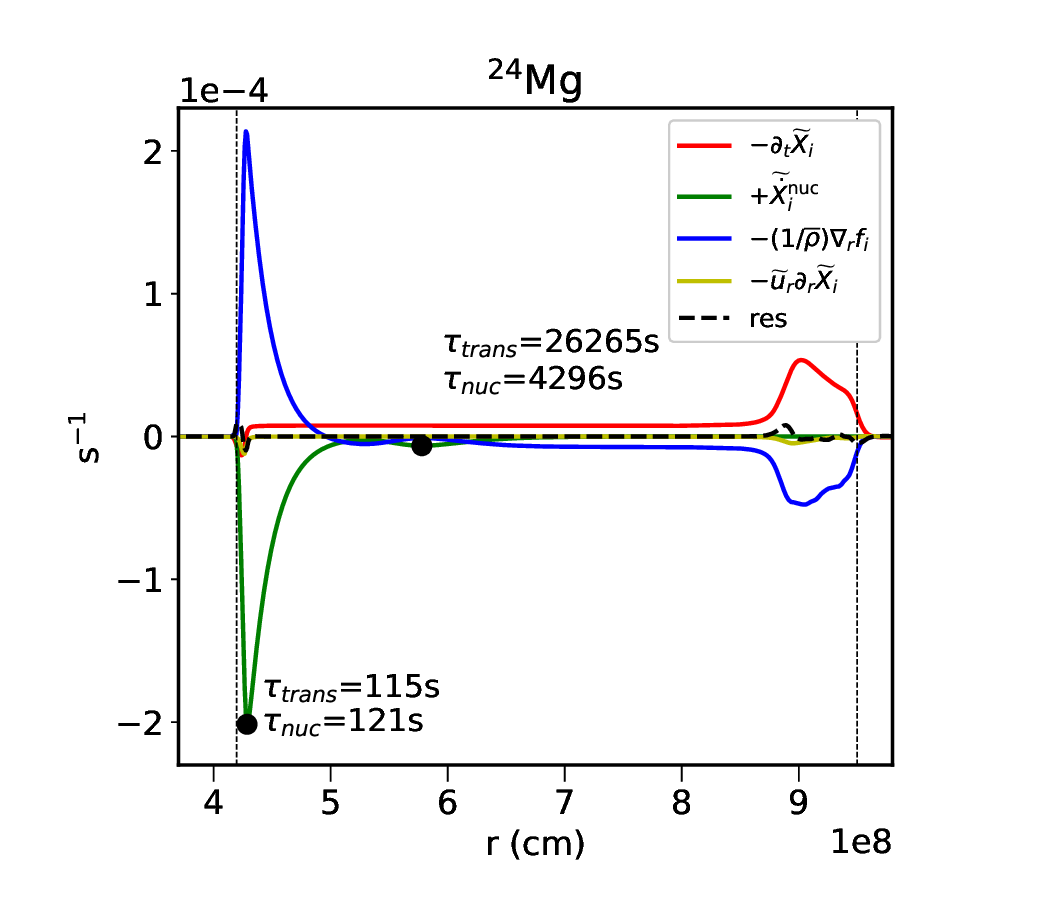}
\includegraphics[width=0.49\hsize]{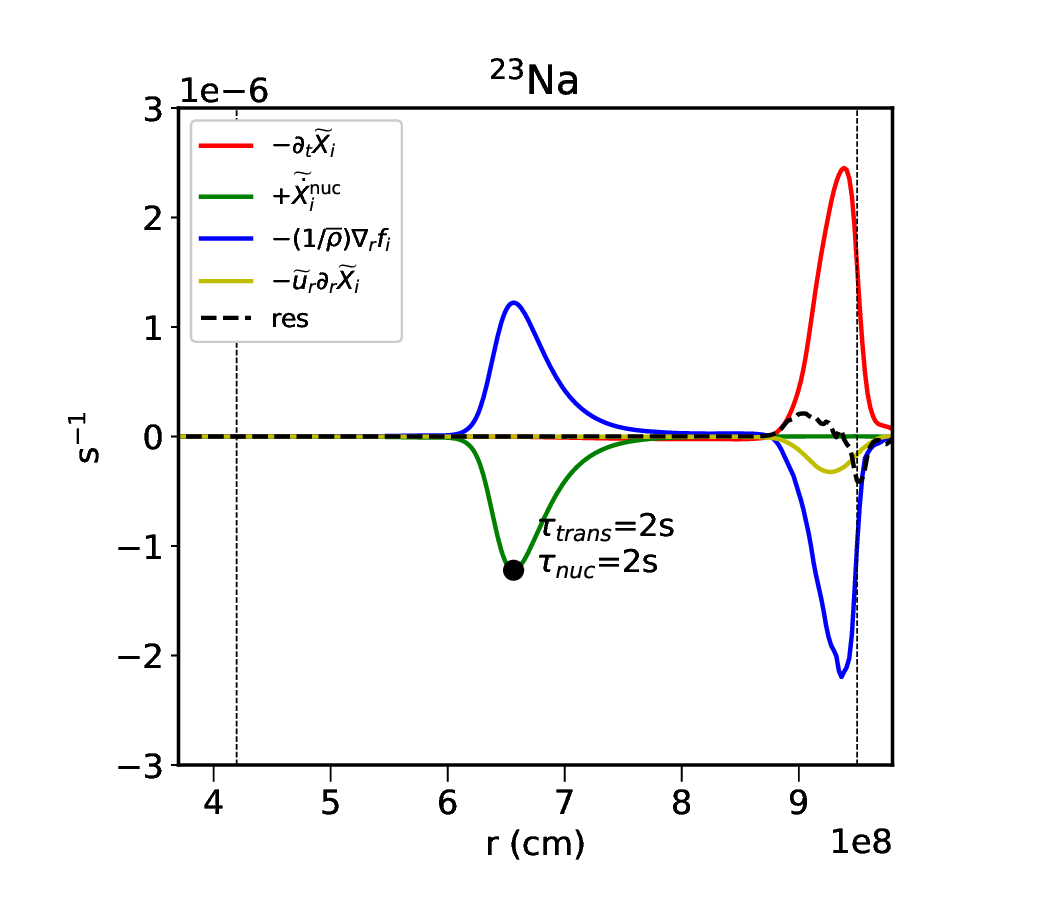}
\caption{Transport Equation for $\elmg{24}$ (left) and $\elna{23}$ (right); spatially and temporally averaged. $\tau_{nuc}$ and $\tau_{trans}$ are nuclear and transport timescales i.e. Eq.\ref{eq:net-nucl-timescale} and Eq.\ref{eq:turbulent-transport-timescale}, respectively.}
\label{fig:ransX_set_mg24_na23}
\end{figure*}

\subsection{$\elmg{24}$}
\label{sect:mg24}

Nuclear burning of $\elmg{24}$ (Fig.\ref{fig:ransX_set_mg24_na23}) is controlled by
Reactions \ref{react:mg24_1}, \ref{react:mg24_2}, \ref{react:mg24_3}, \ref{react:mg24_4}:

\begin{align}
\elhe{4} + \elne{20} \longleftrightarrow & \ \elmg{24} \label{react:mg24_1} \\
 \elh{1} + \elna{23} \longleftrightarrow & \ \elmg{24} \label{react:mg24_2} \\
 \elhe{4} + \elmg{24} \longleftrightarrow & \ \elsi{28} \label{react:mg24_3} \\
 \elc{12} + \elo{16} \longrightarrow & \ \elhe{4} + \elmg{24} \label{react:mg24_4}
\end{align}

We find that $\elmg{24}$ is mainly destroyed ($\fav{\dot{X}}_i^{\rm nuc}$ < 0) in two layers. One is at around the temperature maximum with nuclear destruction peaking at $4.2 \times 10^8$ cm and the other (relatively weak) at $5.8 \times 10^8$~cm. The nuclear timescales there are about $121$~s and $4296$~s, respectively. Whereas the transport timescale for the first peak $115$~s is comparable to the nuclear timescale, it is $26265$~s i.e. 6 times longer at the second peak indicating weak nuclear destruction in a region dominated by transport. Regardless of the entrainment of $\elmg{24}$ from the outer stable layer, where $\partial_t \fav{X}_i$ is comparable to $-(1/\erho)\nabla_r f_i$ from beyond $9 \times 10^8$~cm, its overall density, as indicated by $-\partial_t \fav{X}_i > 0$, decreases throughout the convection zone due to the overall faster nuclear burning. Nevertheless, due to the initial shell merger in 3D and resulting pollution from the upper $\elmg{24}$ rich layer, the mass fraction of $\elmg{24}$ increased in convection zone. Hence, we classify the structure as Type II*+ (Tab.~\ref{tab:structure_summary}).

\subsection{$\elna{23}$}
\label{sect:na23}

Nuclear burning of $\elna{23}$ (Fig.~\ref{fig:ransX_set_mg24_na23}) is controlled by
Reactions \ref{react:na23_1}, \ref{react:na23_2}, \ref{react:na23_3}:

\begin{align}
\elh{1} + \elna{23}  \longleftrightarrow & \ \elmg{24} \label{react:na23_1} \\
\elh{1} + \elna{23} \longleftrightarrow & \ \elhe{4} + \elne{20} \label{react:na23_2} \\
\elc{12} + \elc{12} \longrightarrow & \ \elh{1} + \elna{23} \label{react:na23_3}
\end{align}

We find only one dominant peak with nuclear destruction ($\fav{\dot{X}}_i^{\rm nuc}$ < 0) at $6.5 \times 10^8$ cm with nuclear timescales around 2 s. The transport timescale at this location is the same, indicating a single convective-reactive layer. The source of $\elna{23}$ for the burning is entrainment and turbulent transport from the upper stable layer, where $\partial_t \fav{X}_i \sim -\nabla_r f_i$. We classify the structure as Type I*+ (Tab.~\ref{tab:structure_summary}).

\begin{figure}
\includegraphics[width=0.99\hsize]{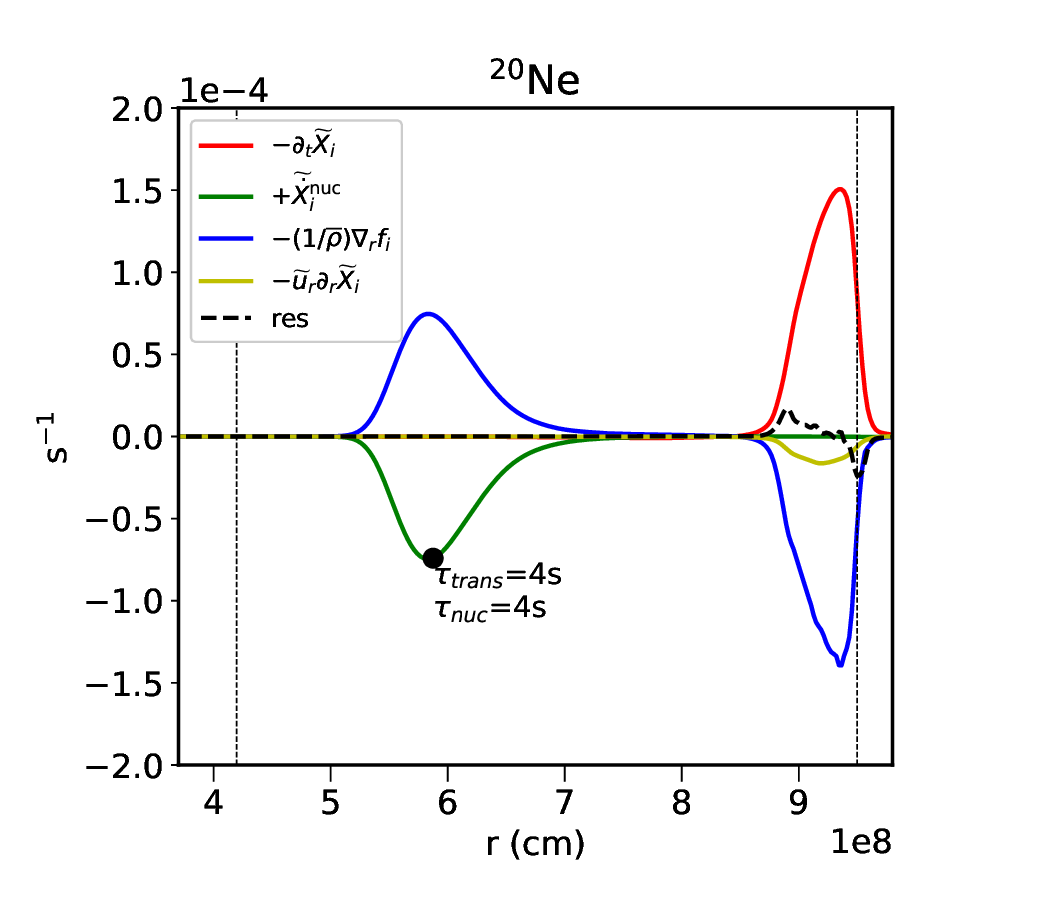} \\

\includegraphics[width=0.99\hsize]{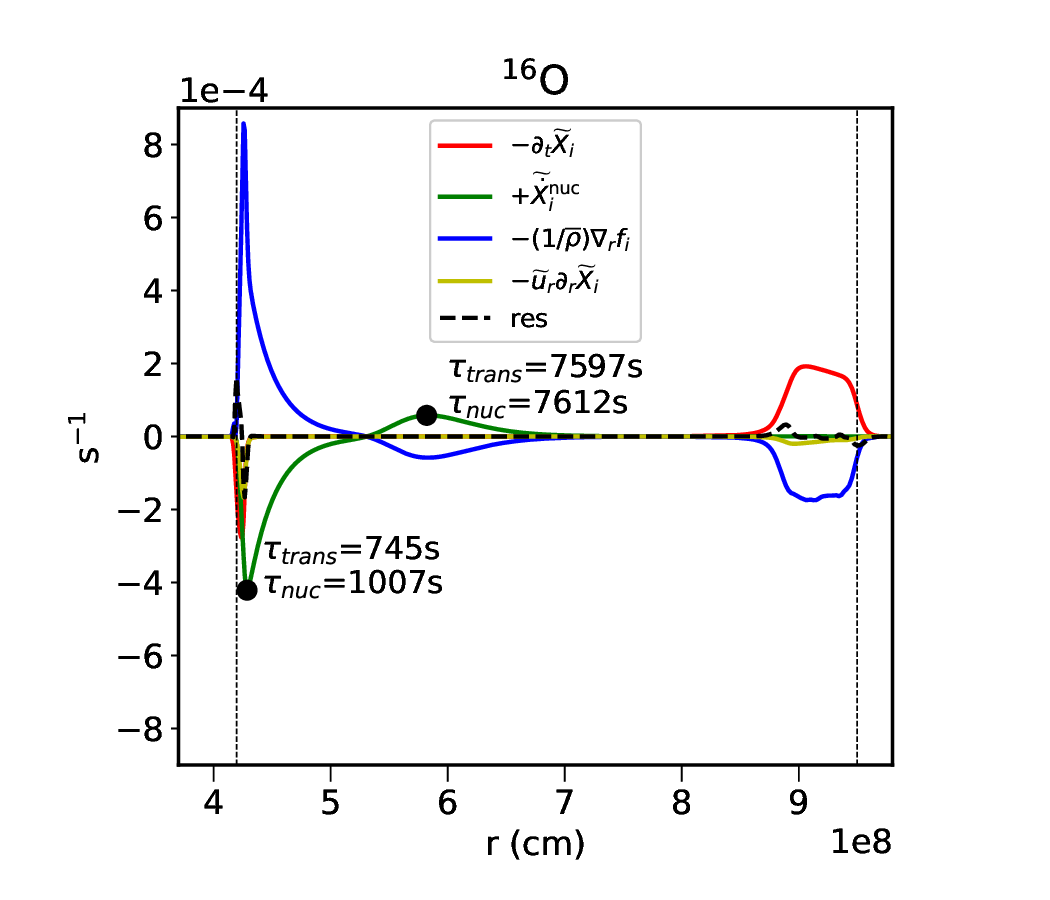} \\

\includegraphics[width=0.99\hsize]{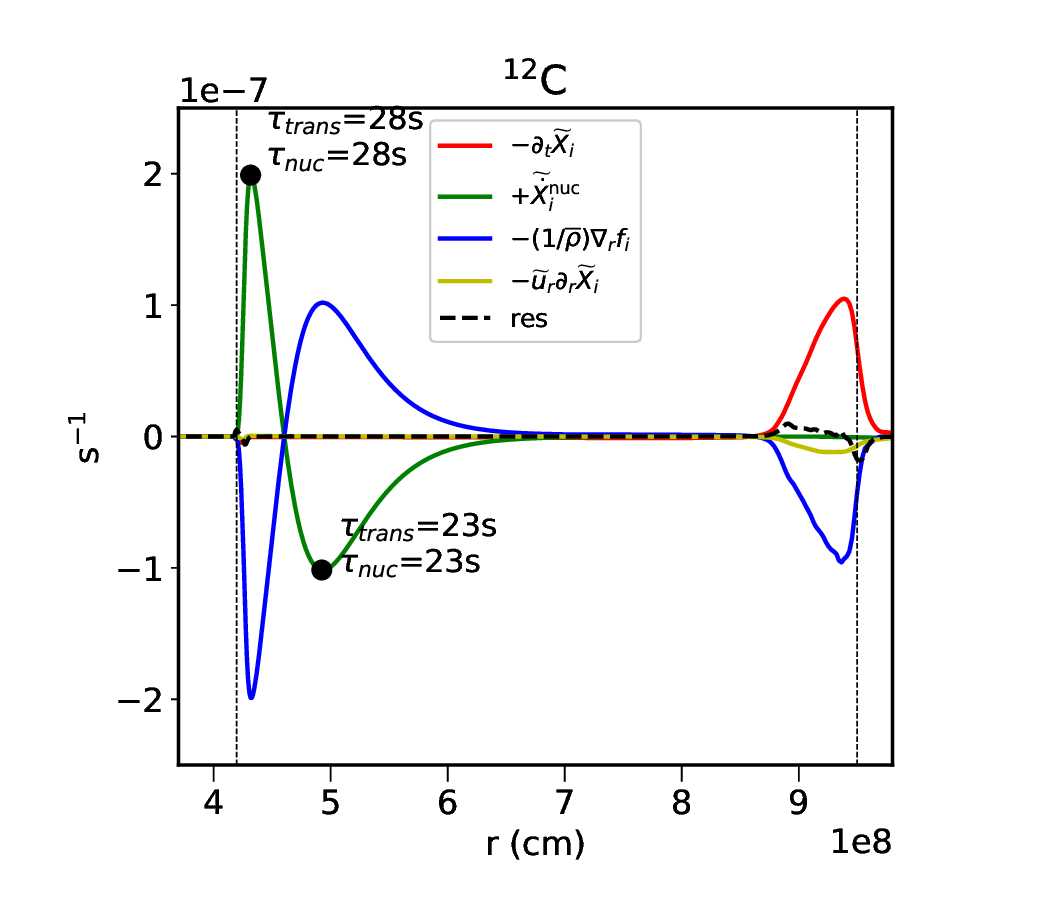}
\caption{Transport Equation for $\elna{20}$ (upper), $\elo{16}$ (middle) and $\elc{12}$ (lower); spatially and temporally averaged. $\tau_{nuc}$ and $\tau_{trans}$ are nuclear and transport timescales i.e. Eq.\ref{eq:net-nucl-timescale} and Eq.\ref{eq:turbulent-transport-timescale}, respectively.}
\label{fig:ransX_set_ne20_o16_c12}
\end{figure}

\begin{figure}
\includegraphics[width=0.99\hsize]{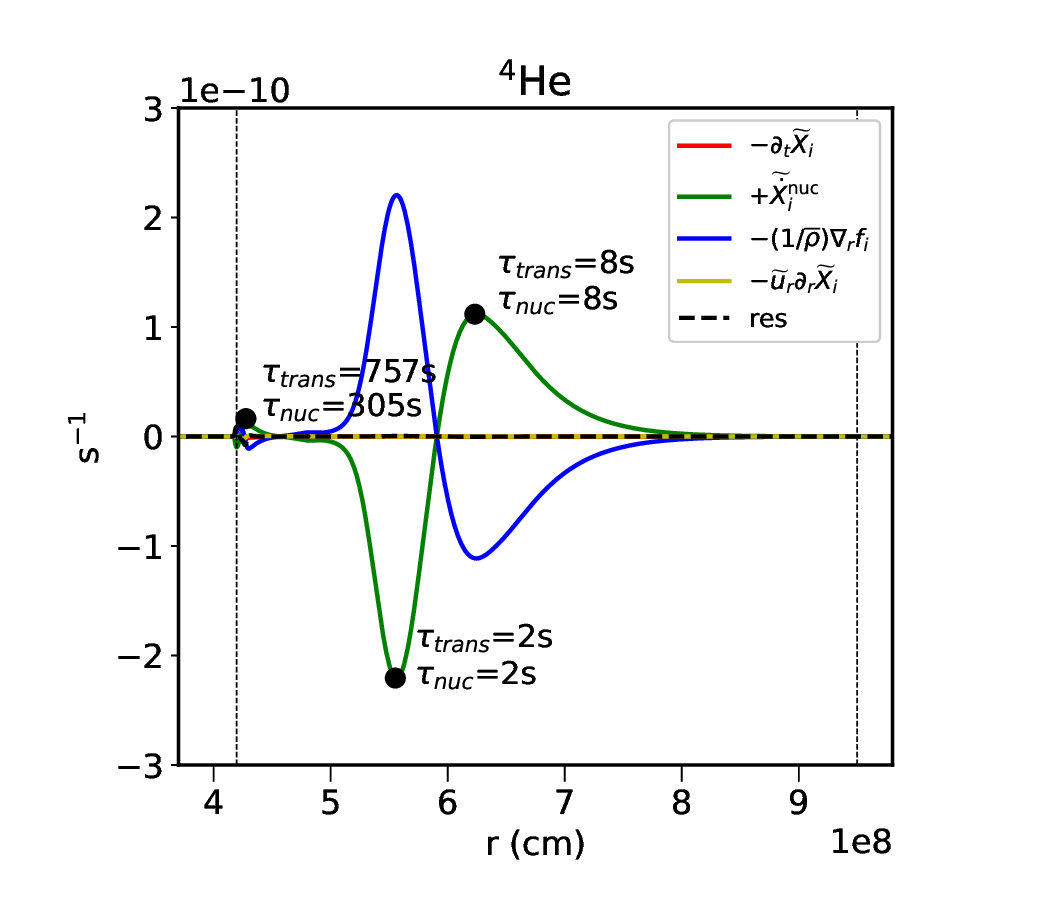} \\

\includegraphics[width=0.99\hsize]{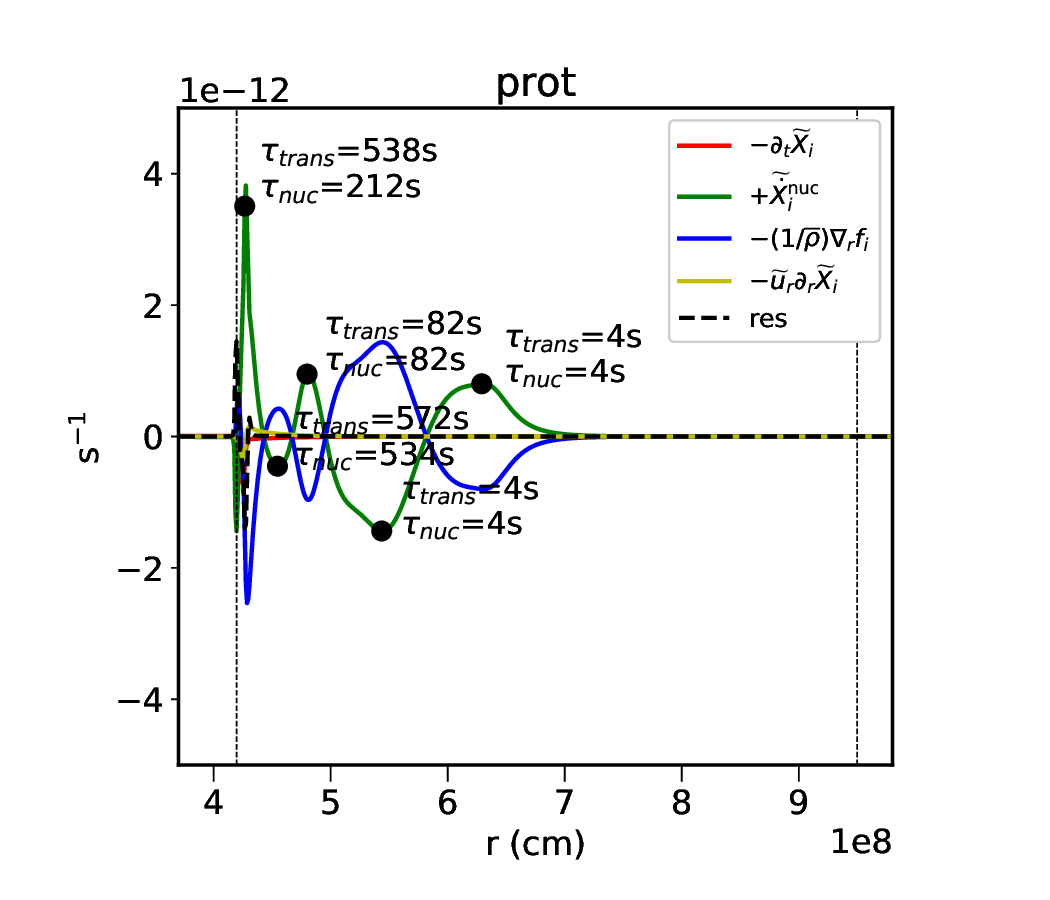} \\

\includegraphics[width=0.99\hsize]{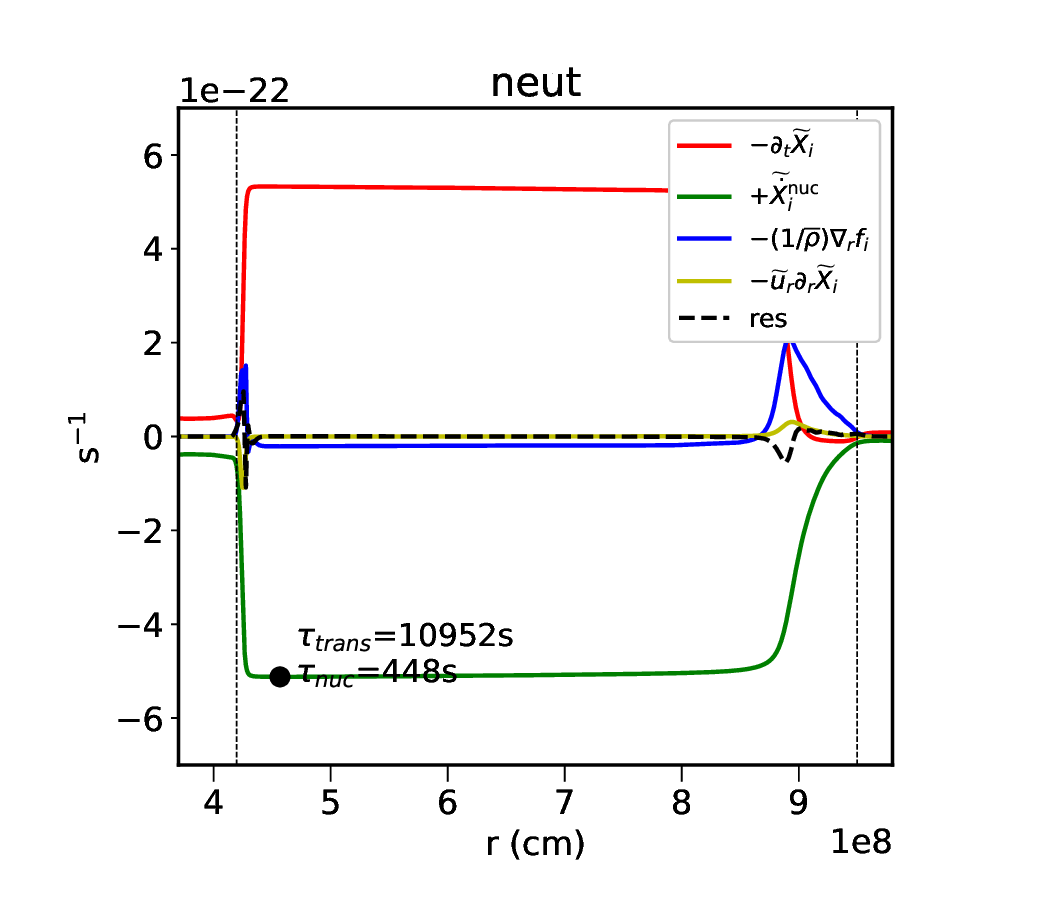}

\caption{Transport Equation for $\elhe{4}$ (upper), $\elh{1}$ (middle) and $neutrons$ (bottom); spatially and temporally averaged. $\tau_{nuc}$ and $\tau_{trans}$ are nuclear and transport timescales i.e. Eq.\ref{eq:net-nucl-timescale} and Eq.\ref{eq:turbulent-transport-timescale}, respectively.}
\label{fig:ransX_set_he4_h1_neut}
\end{figure}

\subsection{$\elne{20}$}
\label{sect:ne20}

Nuclear burning of $\elne{20}$ (Fig.\ref{fig:ransX_set_ne20_o16_c12}) is controlled by Reactions \ref{react:ne20_1}, \ref{react:ne20_2}, \ref{react:ne20_3},\ref{react:ne20_4}, \ref{react:ne20_5}, \ref{react:ne20_6}:

\begin{align}
\elhe{4} + \elo{16}  \longleftrightarrow & \ \elne{20}  \label{react:ne20_1} \\
\elhe{4} + \elne{20}  \longleftrightarrow & \ \elmg{24} \label{react:ne20_2}  \\
\elh{1} + \elna{23} \longleftrightarrow & \ \elhe{4} + \elne{20} \label{react:ne20_3}  \\
 \elc{12} + \elne{20} \longleftrightarrow & \ \elhe{4} + \elsi{28} \label{react:ne20_4}  \\
 \elc{12} + \elc{12} \longrightarrow & \ \elhe{4} + \elne{20} \label{react:ne20_5}  \\
\elh{1} + \elna{23} \longrightarrow & \ \elc{12} + \elne{20} \label{react:ne20_6}
\end{align}

We find a nuclear destruction region ($\fav{\dot{X}}_i^{\rm nuc}$ < 0) with a peak at $5.8 \times 10^8$ cm balanced by transport with timescales around 4 s, indicating a single convective-reactive layer. At temperatures T > $10^9$ K many nuclei undergo photodisintegration. The first major disintegration phase is the $\elne{20} + \gamma \rightarrow \ \elo{16} + \elhe{4}$ (React. \ref{react:ne20_1}). At about $T \sim 1.3 \times 10^9$ K, this rate becomes greater than the rate for $\elo{16} + \elhe{4} \rightarrow \ \elne{20} + \gamma$ (React. \ref{react:o16_2}), whereupon the $\elne{20}$ is effectively decayed. The released $\alpha$ particles are captured by the remaining $\elne{20}$, so the net effect is

\begin{align}
2 \ \elne{20} \rightarrow \ \elo{16} + \elmg{24}
\label{react:ne20_to_o16_mg24}
\end{align}

While this picture seems to be consistent with the $\elne{20}$ destruction peak correlating with the $\elo{16}$ production peak (to be described in the next section.\ref{sect:o16}), this is not the case for $\elmg{24}$, which is also destroyed at this location. This indicates that the evolution of $\elmg{24}$ is only weakly affected by the burning of $\elne{20}$. The source of $\elne{20}$ for the burning is the entrainment and turbulent transport from the upper stable layer, where $\partial_t \fav{X}_i \sim -(1/\erho)\nabla_r f_i$. We classify the structure as Type I*+ (Tab.~\ref{tab:structure_summary}).

\subsection{$\elo{16}$}
\label{sect:o16}

Nuclear burning of $\elo{16}$ (Fig.\ref{fig:ransX_set_ne20_o16_c12}) is controlled by
Reactions \ref{react:o16_1}, \ref{react:o16_2}, \ref{react:o16_3},
\ref{react:o16_4}, \ref{react:o16_5}:

\begin{align}
\elhe{4} + \elc{12} \longleftrightarrow & \ \elo{16} \label{react:o16_1} \\
\elhe{4} + \elo{16} \longleftrightarrow & \ \elne{20} \label{react:o16_2}  \\
\elc{12} + \elo{16}  \longrightarrow & \ \elhe{4} + \elmg{24} \label{react:o16_3}  \\
 \elo{16} + \elo{16} \longrightarrow & \ \elh{1} + \elp{31} \label{react:o16_4}  \\
 \elo{16} + \elo{16} \longrightarrow & \ \elhe{4} + \elsi{28} \label{react:o16_5}
\end{align}

We find a nuclear destruction layer ($\fav{\dot{X}}_i^{\rm nuc}$ < 0) with a peak at a temperature maximum around $4.2 \times 10^8$ cm and a nuclear timescale around $1007$ s. We also find a nuclear production region ($\fav{\dot{X}}_i^{\rm nuc}$ > 0) with a peak around $5.8 \times 10^8$ cm and a nuclear timescale around $7612$ s, which correlates with the $\elne{20}$ destruction peak indicating that its source is $\elne{20}$ destruction due to photodisintegration (React.\ref{react:ne20_to_o16_mg24}). Another source of $\elo{16}$ for the burning is the entrainment and turbulent transport from the upper stable layer, where $\partial_t \fav{X}_i \sim -(1/\erho)\nabla_r f_i$. We classify the structure as Type II*+ (Tab.~\ref{tab:structure_summary}).

\subsection{$\elc{12}$}
\label{sect:c12}

Nuclear burning of $\elc{12}$ (Fig.\ref{fig:ransX_set_ne20_o16_c12}) is controlled by the following
Reactions \ref{react:c12_1}, \ref{react:c12_2}, \ref{react:c12_3},
\ref{react:c12_4}, \ref{react:c12_5}, \ref{react:c12_6}, \ref{react:c12_7}:

\begin{align}
\elhe{4} + \elc{12} \longleftrightarrow & \ \elo{16} \label{react:c12_1} \\
\elc{12} + \elne{20}  \longleftrightarrow & \ \elh{1} + \elp{31} \label{react:c12_2} \\
\elc{12} + \elne{20} \longleftrightarrow & \ \elhe{4} + \elsi{28} \label{react:c12_3} \\
\elc{12} + \elc{12} \longrightarrow & \ \elh{1} + \elna{23} \label{react:c12_4} \\
\elc{12} + \elc{12} \longrightarrow & \ \elhe{4} + \elne{20} \label{react:c12_5} \\
\elc{12} + \elo{16} \longrightarrow & \ \elhe{4} + \elmg{24} \label{react:c12_6} \\
\elc{12} \longleftrightarrow & \ \elhe{4} + \elhe{4} + \elhe{4} \label{react:c12_7}
\end{align}

We find one nuclear production region ($\fav{\dot{X}}_i^{\rm nuc}$ > 0) with peak at around $4.2 \times 10^8$ cm just above temperature maximum and one nuclear destruction region ($\fav{\dot{X}}_i^{\rm nuc}$ < 0) with peak around $5 \times 10^8$~cm. The nuclear and transport timescales are at those locations identical 28~s and 23~s, respectively indicating double convective-reactive structure. The apparent $\elc{12}$ production is perhaps dominated by the reverse of the triple-$\alpha$ channel (React.~\ref{react:c12_7}); however, a definitive identification of the dominant contributor requires reaction-by-reaction diagnostics and we flag this for follow-up. Besides the nuclear production region, another source of $\elc{12}$ for the burning is the entrainment and turbulent transport from the upper stable layer, where $\partial_t \fav{X}_i \sim -(1/\erho)\nabla_r f_i$. We classify the structure as Type II*+ (Tab.~\ref{tab:structure_summary}).

\subsection{$\elhe{4}$ and $\elh{1}$}
\label{sect:he4_h1}

$\elhe{4}$ and $\elh{1}$ burning is already encompassed in reactions from previous
sections (Sec.~\ref{sect:cl35} -- \ref{sect:si28} and Sec.~\ref{sect:na23} -- \ref{sect:c12}).

For $\elhe{4}$ (Fig.~\ref{fig:ransX_set_he4_h1_neut}) we identify a major nuclear production region ($\fav{\dot{X}}_i^{\rm nuc}$ > 0) with peak at $6.2 \times 10^8$ cm and a nuclear destruction region ($\fav{\dot{X}}_i^{\rm nuc}$ < 0) peaking at about $5.5 \times 10^8$ cm with nuclear timescales equalling the transport timescales at around 8~s and 2~s, respectively. This suggests a double convective-reactive structure associated with mutual redistribution of helium between them. Thus, whatever $\elhe{4}$ is produced in the second/outer layer is burned in the first/inner layer. We also identify a minor nuclear production region of $\elhe{4}$ at temperature maximum around $4.2 \times 10^8$ cm with transport and nuclear timescales being 757~s and 305~s, respectively. We classify the structure as Type III+ (Tab.~\ref{tab:structure_summary}).

\begin{table*}
\caption{Summary of convective structures for each of the isotopes. Structure Type is explained in Fig.\ref{fig:type_explainer}, Layers is number of layers, $D_a^{peak}$ is Damk\"ohler number for individual layers at the peak of nuclear burning in each layer as shown in Figs.\ref{fig:ransX_set_ar36_cl35}-\ref{fig:ransX_set_he4_h1_neut} ($D_a = \tau_{trans}/ \tau_{nuc}$), Layer 1-5 is position identifier of the layers from $T_{max}$ to the right, log$_{10} (\fav{X}_i/\fav{X}_{i}^{ini})$ is log$_{10}$ of ratio of mass fraction $\fav{X_i}$ and initial $\fav{X}_i^{ini}$ of isotope $i$ in the middle of convection zone ($6.85 \times 10^8$ cm) at the central time $t_c = 2221 s$. 
}
\label{tab:structure_summary}
\begin{tabular}{|c|c|c|c|c|c|c|c|c|}
\hline
Isotope & Structure Type & Layers & \multicolumn{5}{c|}{$D_a^{peak}$} & log$_{10} (\fav{X}_i/\fav{X}_{i}^{ini})$  \\
\cline{4-8}
 &  &  & Layer1 & Layer2 & Layer3 & Layer4 & Layer5 &  \\
\hline
$neutrons$      & Type I+        & 1 & 24.4 & - & - & - & - & $+1.9$ \\
$protons$      & Type V+        & 5 & 2.5 & 1.1 & 1.0 & 1.0 & 1.0 & $+2.4$ \\
$\elhe{4}$    & Type III+      & 3 & 2.5 & 1.0 & 1.0  & - & - & $+3.1$ \\
$\elc{12}$  & Type II$^{*}$+ & 2 & 1.0 & 1.0 & - & - & - & $+1.7$ \\
$\elo{16}$  & Type II$^{*}$+ & 2 & 0.7 & 1.0 & - & - & - &  $+1.4 \times 10^{-1}$\\
$\elne{20}$ & Type I$^{*}$+  & 1 & 1.0 & - & - & - & - & $+3.2$ \\
$\elna{23}$ & Type I$^{*}$+  & 1 & 1.0 & - & - & - & - &  $+4.6$\\
$\elmg{24}$ & Type II$^{*}$+ & 1 & 1.0 & 6.1 & - & - & - &  $+2.8 \times 10^{-2}$\\
$\elsi{28}$ & Type II$^{*}$- & 2 & 0.8 & 1.1 & - & - & - &  $-9.4 \times 10^{-2}$\\
$\elp{31}$  & Type II-       & 2 & 1.0 & 1.0 & - & - & - & $ -7 \times 10^{-1}$\\
$\els{32}$  & Type I$^{*}$- & 1 & 1.4 & 1.1 & - & - & - &  $-1.7 \times 10^{-1}$ \\
$\els{34}$ & Type II+       & 2 & 1.0 & 1.0 & - & - & - &  $+3.5$ \\
$\elcl{35}$ & Type III+      & 3 & 1.0 & 1.0 & 1.0 & - & - &  $+8.3 \times 10^{-1} $\\
\hline
\end{tabular}
\end{table*}

For the $\elh{1}$/$protons$ (Fig.\ref{fig:ransX_set_he4_h1_neut}) we find three nuclear production regions ($\fav{\dot{X}}_i^{\rm nuc}$ > 0) with peaks at radii of $4.2 \times 10^8$, $4.6 \times 10^8$ cm and $6.2 \times 10^8$ cm and two nuclear destruction regions ($\fav{\dot{X}}_i^{\rm nuc}$ < 0) with peaks at radii of $4.5 \times 10^8$ cm and $5.3 \times 10^8$ cm. The transport timescales closely match the nuclear timescales, except the first peak, where transport timescale 538 s is almost 2x longer than nuclear timescale 221 s. This indicates a quintuple convective-reactive layer, with mutual redistribution and burning of $\elh{1}$ between these layers. We classify the structure as Type V+ (Tab.~\ref{tab:structure_summary}).

\subsection{Neutrons}
\label{sect:neut}

Neutrons (Fig.~\ref{fig:ransX_set_he4_h1_neut}) just decay as shown by Reaction \ref{react:neut_1}. We nevertheless track $neutrons$ explicitly to close the reduced network consistently and to capture neutron creation/decay bookkeeping in the reaction rate evaluation even when neutron capture channels are not included.

\begin{align}
\elh{1} \longleftarrow & \ neutrons  \label{react:neut_1}
\end{align}

Whereas the nuclear timescale is around 448~s for this decay, the transport timescale is much longer at around $10952$~s. This leads to steady decay of neutrons in the convection zone, where turbulent transport plays only minor role. Compared to the initial composition, we observe an increase in neutron abundance in the middle of convection zone due to the initial merging event that effectively smoothed out the initial neutron composition gradient. Therefore we classify the structure of Type I+ (Tab.~\ref{tab:structure_summary}). It is unclear what an extended nuclear network with additional n-capture or n-production reactions would do to the reported neutron structure, hence rendering conclusions for neutrons as preliminary. In any case, neutron impact on $\elh{1}$ would be negligible as its abundance is lower by 9 and 2 orders of magnitude at the bottom and top of convection zone, respectively (Fig.~\ref{fig:X_evolved}).

\subsection{Summary of quasi-stable convection zone layering}
\label{sect:summ_structures}

In summary, we find a complex layering within the contiguous convection zone. This is caused by the competition between turbulent transport (including entrainment at the convective boundaries) and nuclear burning. Table~\ref{tab:structure_summary} summarises these layers, giving our Type~I--V classification for each species, as well as the Damk\"ohler number for every layer. As described in Sec.~\ref{sec:transport_equation}, our classification scheme is based on the spatial structure of the transport balance (i.e. on the sign and relative magnitude of $\partial_t\fav{X}_i$, $\fav{\dot{X}}_i^{\rm nuc}$ and $-(1/\erho)\nabla_r f_i$), rather than on the abundance profile $\fav{X}_i$ alone. We discuss this distinction more below.

The types of structures detected within the simulated convection zone can be summarized as follows:

\begin{itemize}
\item Single convective layer, where isotope density is mainly controlled by nuclear reactions and turbulent transport plays a minor role (eg. $neutrons$)
\item Single convective reactive layer, where the total amount of isotopes destroyed by
nuclear burning is equal to the amount of material entrained from the upper stable layer and redistributed by turbulence ($\elne{20}, \elna{23}$)
\item Single convective-reactive layer with two nuclear burning peaks, where the total amount of isotopes produced by nuclear burning becomes a major contributor to the upper stable layers ($\els{32}$)
\item Double convective-reactive layer, where one layer produces specific isotopes and becomes a fuel source for the other by transport ($\els{34}, \elp{31}, \elc{12}, \elhe{4}$).
\item Triple convective-reactive layer, where nuclear production in two layers becomes the source for the third ($\elcl{35}$)
\item Quintuple convective-reactive layer, where nuclear production from three layers via turbulent transport acts as a source for nuclear burning in the remaining two ($\elh{1}$)
\end{itemize}

In all cases, the spatial domains of reaction channels overlap, impacting the number of distinct effective layers. Conversely, sharp entrainment at boundaries can accentuate layering by supplying contrasting composition on mixing times shorter than or comparable to burning times (eg. $\elne{20}$).

Other observed structures are just modifications of these six. For example, the structure of
$\elsi{28}$ is essentially a double convective-reactive layer with some transport from inner to outer stable layers unaffected by burning.  
Although entrainment of $\elmg{24}$ from the upper boundary tends to increase its density inside the convection zone, its nuclear burning destruction within leads to exactly the opposite situation and its density decreases everywhere ($\partial_t \fav{X}_i < 0$). 

In fact, when abundance just above or below the convective zone for a given isotope is lower than within it and the convective boundary is weaker there (compared to potential higher abundance at the opposite side), the entrainment acts to dilute the convective zone composition of that isotope.  At the same time, when the species is being produced within the convection zone, so that production and dilution operate in competition, the net abundance evolution will reflect the balance between the two (e.g. $\els{32},\elsi{28}$). This behaviour contrasts with other cases, in which the entrained abundances exceed those of the convection zone and entrainment instead enriches the convective region ($\elmg{24},\elna{23},\elna{20},\elo{16},\elc{12}, neut$). For the remaining isotopes in our analysis ($\elcl{35},\els{34},\elp{31},\elhe{4},\elh{1}$), this behaviour seems not to be prominent due to $\partial_t \fav{X}_i$ at convection boundaries being negligible compared to nuclear burning $\fav{\dot{X}}_i^{\rm nuc}$ or transport $-(1/\erho)\nabla_r f_i$) within the convection zone.

Table $\ref{tab:structure_summary}$ shows that the Damk\"ohler number $\mathrm{Da}$ at peak nuclear burning is close to unity for most layers and species. This behaviour can be understood as a consequence of nuclear burning readily matching the rate at which material is transported into the burning region. As a species is mixed inward to progressively deeper layers, the timescale for its destruction (or production) becomes very short, because the burning timescale depends strongly on temperature, density, and composition, and generally decreases with depth. The precise location of a given burning layer is set by the efficiency of convective/turbulent transport: slower transport places the layer further out, while faster transport shifts it inward. In all such cases the inflow rate can be balanced by burning, so that $\mathrm{Da} \simeq 1.0$. Furthermore, in 3D, stochastic fluctuations produce a spectrum of transport speeds, which broadens the burning shell in temporally and spatially averaged (e.g. RANS) profiles. In this picture, nuclear burning is the primary agent establishing compositional layering through the reaction rates, while turbulence subsequently smooths these layers. We note, however, that the mixing rate still sets the exact location of each layer. An important exception occurs at the base of the convection zone. Here, material cannot be transported to greater depths and therefore cannot access the correspondingly shorter burning timescales. In this region, burning may be unable to keep pace with the inflow, resulting in $\mathrm{Da} < 1.0$. We attribute the sub-unity values observed for $\elo{16}$ and $\elsi{28}$ (see Figs.\ref{fig:ransX_set_p31_si28},\ref{fig:ransX_set_ne20_o16_c12}) to this effect. Conversely, most of the cases with $\mathrm{Da} > 1.0$ in Table $\ref{tab:structure_summary}$ are subject to caveats — very small compositional gradients or proximity to the turbulent base of the convection zone. The one exception is $\els{32}$, whose transport profile differs markedly from those of the other species. In this case the species is both abundant and net-produced within the convection zone, while simultaneously being diluted by entrainment, which may account for its anomalous Damk\"ohler number. 

\subsection{Robustness of observed layering}

Here we address the temporal robustness of our results. After an initial transient, during which the abrupt features of the initial $\fav{X}_i$ profiles are smoothed out (Fig.~\ref{fig:X_evolved}), the RANS profiles settle into a quasi-steady state in which the three terms of Eq.~\ref{eq:transport-comp2} balance each other in a self-regulating manner. The fact that $\mathrm{Da} \simeq 1.0$ for most species (Table~\ref{tab:structure_summary}) is itself an indication of this stability: burning readily adjusts to the rate at which material is transported into the burning region, so that the radial position and number of the reactive layers remain essentially fixed over the averaging interval. Consequently, the Type~I--V classification does not change during the analysed period, since the layering is primarily imprinted by the reaction rates through the temperature and density stratification, with turbulence acting only to smooth and slightly displace the layers. Some sensitivity is nevertheless expected for the marginal cases, i.e. those with very small compositional gradients, those located close to the turbulent base of the convection zone, and $\els{32}$, where production and entrainment-driven dilution compete; here the balance may shift and the assigned type is correspondingly less robust. Whether the classification remains unchanged on timescales considerably longer than those simulated here cannot be assessed with the present data and requires longer simulations.

\section{Discussion}
\label{sect:discussion}

\subsection{1D Stellar Evolution Theory Implications}

\subsubsection{Signatures of layering in abundance profiles?}

As noted above, our analysis is based on the spatial structure of the transport balance (via the transport equation, see Sec.~\ref{sec:transport_equation}), rather than on the abundance profile $\fav{X}_i$. An interesting question to ask is if the detected layering is apparent in the chemical abundance profiles. First we note that the abrupt features in the initial $\fav{X}_i$ profiles are clearly smoothed in the evolved $\fav{X}_i$ (Fig.~\ref{fig:X_evolved}), through the turbulent mixing (and burning). It can be seen that the layering in the transport equation analysis is not easily detectable by looking at the abundance profiles (Fig.~\ref{fig:X_evolved}). The reason for this is related to our findings that (i) most of the layers have $Da \sim 1$ (Tab.~\ref{tab:structure_summary}), which leads to a steady-state/constant abundance (at least while fuel can be replenished through mixing), and (ii) the smoothing-out effect of turbulence, due to its spectrum of mixing and multidimensional nature (Appendix \ref{ap:snapshots}) -- something which is rarely taken into account in 1D stellar structure models.

\subsubsection{Fundamental Challenges to Mixing Length Theory}

Our results highlight regimes where burning and turbulent transport interact on comparable timescales (Damk\"ohler number near unity), challenging the common assumption of instantaneous mixing. However, 1D stellar evolution codes do include nuclear burning and, in some cases, time-dependent mixing schemes that can, with sufficiently small timesteps, approximate some aspects of convective-reactive evolution \citep{Jermyn_2023} or mergers. We therefore view our findings as complementary: they could help to identify where simple prescriptions like diffusion approximations are likely adequate and where they are not. This work could guide us on how to constrain interaction between transport and burning within the physics of 1D stellar evolution codes by derivation of appropriate models for composition flux $f_i$ impacted by nuclear burning within turbulent flow.

A practical difference between 1D and 3D can also arise from numerical time stepping: in 1D, timesteps that are large compared to local nuclear and transport timescales can smear or miss transient convective-reactive balances, even when composition gradients are present (Fig.~\ref{fig:X_evolved}). Our Type~I--V classification is therefore best viewed as a diagnostic of the local balance of burning and transport, which could be recovered in 1D only if both the mixing prescription and the timestep resolution capture the relevant timescales.
We note that Damk\"ohler numbers larger than unity (transport slower than burning) have been reported in related 3D shell interaction studies; see, e.g., \citet{Rizzuti2024}.

\subsubsection{Inhomogeneous Chemical Mixing and Nucleosynthesis}
Different isotopes in our simulations exhibit distinct layered structures (Types I-V as classified in the paper) due to inclusion of more complete turbulence physics i.e. something that no diffusion based approximation can yield. It implies that stellar yields may be different from homogeneous mixing models even when enhanced by a diffusive mixing approximation. Heterogeneity in chemical distribution altered by turbulence during pre-supernova evolution could open new pathways beyond traditional explosive nucleosynthesis scenarios and lead to new abundance patterns in supernova ejecta, affecting both local and global chemical enrichment patterns.

It has also been shown that the unusual conditions in convective-reactive events could lead to understanding of the long-standing underproduction problem for odd-Z elements (P, Cl, K, Sc) in galactic chemical evolution models (see e.g. \citealt{issa2025,Ritter2017}). Also the identification of specific convective-reactive layers that can efficiently alter production of these elements through \(\alpha\)-capture and proton-induced reactions could even provide a solution to abundance patterns observed in metal-poor stars due to pollution from Pop III supernovae \citep{Umeda2005}.  

\subsubsection{Supernova Progenitor Structure Modifications}
The complex convective-reactive structures identified in the paper alter the composition gradients in pre-supernova models (Fig.\ref{fig:X_evolved}). These structural changes directly impact the properties of the Si/O interface which is a predictor of supernova explodability \citep{boccioli2023}. 

\section{Conclusion}
\label{sect:conclusion}

This study has presented a comprehensive analysis of the fine structure of convective-reactive zones in a massive pre-supernova star through three-dimensional hydrodynamic simulations and Reynolds-Averaged Navier-Stokes (RANS) mean-field analysis. Our investigation of an oxygen-neon shell merger has revealed fundamental insights that challenge current understanding of stellar evolution and nucleosynthesis.

Our primary findings demonstrate that convective-reactive mixing in stars is non-instantaneous, but the resulting ``fine structure'' is not always most apparent in the horizontally averaged abundance profiles $\fav{X}_i$ (Fig.~\ref{fig:X_evolved}), which can appear comparatively flat for many species in 3D. Instead, the layered behavior is most clearly diagnosed from the transport balance in Eq.~\ref{eq:transport-comp2}: distinct regions where nuclear processing is balanced by turbulent transport (and, in some cases, by local time-dependence and entrainment at boundaries). Based on this transport balance we classify composition fine structures into Types~I--V.

The emergence of these structures is primarily driven by finite mixing timescales, where nuclear burning and turbulent transport operate on comparable timescales (Damk\"ohler number $D_a \approx 1$). This work reveals regimes where the interaction between nuclear physics and fluid dynamics cannot be decoupled.

We have identified six distinct categories of convective-reactive behaviour: (1) single convective layers dominated by nuclear reactions, (2) single convective-reactive layers with balanced entrainment and burning, (3) double convective-reactive layers with transport-mediated fuel exchange, (4) double layers producing material for upper stable regions, (5) triple layers with dual production sources, and (6) quintuple layers exhibiting complex multi-scale interactions. Each category exhibits unique nuclear timescales ranging from seconds to thousands of seconds, with transport timescales that may differ by orders of magnitude from convective turnover times.

This work provides further motivation for incorporating full 3D hydrodynamic results into 1D stellar evolution calculations, requiring the development of new computational frameworks that can handle the coupling between nuclear networks, transport equations, and convective flows. The RANS approach demonstrated here provides a pathway for bridging 3D simulations with 1D stellar evolution codes while maintaining computational feasibility. 

The complex interplay between turbulent convection, nuclear burning, and chemical transport identified in this study necessitates the ongoing development of comprehensive multi-physics simulation codes that can simultaneously handle radiation transport, hydrodynamics, nuclear networks, and turbulent mixing processes including compressibility. The domain of applicability where such rigor is required may be narrow, but such codes will be essential for predicting the full range of observable consequences revolving around convective-reactive stellar evolution. 

\section*{Acknowledgements}

MM used generative artificial intelligence (AI) tools solely for assistance with language editing, grammar correction, and improvements to the clarity of the manuscript text. VV acknowledges support from the Royal Astronomical Society
Research Fellowship. RH acknowledges support from STFC, the
World Premier International Research Centre Initiative (WPI Initiative), MEXT, Japan, the IReNA AccelNet Network of Networks
(NSF, Grant No. OISE-1927130), CeNAM (grant DE-SC0026204)
and the Wolfson Foundation that part-funded the greenHPC facility at Keele. This work used the DiRAC Memory Intensive service
(Cosma) at Durham University, managed by the Institute for Computational Cosmology on behalf of the STFC DiRAC HPC Facility
(www.dirac.ac.uk). The DiRAC service at Durham was funded by
BEIS, UKRI and STFC capital funding, Durham University and
STFC operations grants. DiRAC is part of the UKRI Digital Research Infrastructure. FR is a fellow of the Alexander von Humboldt
Foundation, and acknowledges support by the Klaus Tschira Foundation and by the INAF Mini grant 2024, ‘GALoMS – Galactic
Archaeology for Low Mass Stars’ (1.05.24.07.02). 
S.W.C. acknowledges federal funding from the Australian Research Council through a Future Fellowship (FT160100046) and Discovery Projects (DP190102431 \& DP210101299). Parts of this research was supported by the Australian Research Council Centre of Excellence for All Sky Astrophysics in 3 Dimensions (ASTRO 3D), through project number CE170100013. This research was supported by use of the Nectar Research Cloud, a collaborative Australian research platform supported by the National Collaborative Research Infrastructure Strategy (NCRIS).

\appendix

\section{Averaging Procedure}
\label{appendix:averaging}

\par We obtain our 1D RANS equations by introducing two types of averaging: horizontal averaging and statistical averaging \citep{Besnard1992}. In practice, statistical averages are computed by performing a time average (the ergodic hypothesis\footnote{As it is not feasible to calculate "infinite" realizations of our simulation to obtain true statistical averages, we take advantage of ergodic hypothesis and in both cases deal with time averages over several convective turnover timescales instead.}). Therefore, the combined average of a quantity $q$ is defined as

\begin{align}\label{eq:eht}
\eht{q}(x,t) = \frac{1}{T\Delta V}\int_{t- T/2}^{t+T/2} q(x,y,z,t')~dVdt'
\end{align}

where $dV$ is volume of grid zone\footnote{For Cartesian geometry is $dV = dxdydz$, for spherical geometry is $dV = 1/3 d r^3 d cos \theta d\phi$},
$T$ is the averaging time period, and $\Delta V$ is total volume being averaged over. This allows for decomposition of any quantity into its mean and fluctuating part. We will denote fluctuations always by quotation marks. Such decomposition looks for example for density $\rho$ as follows: $\rho = \eht{\rho} + \rho'$, where $\rho = \rho(x,y,z), \rho' = \rho'(x,y,z)$ and
$\eht{\rho} = \eht{\rho}(x)$.

In case of compressible turbulence \citep{Canuto1997}, it is appropriate to express mean fields using mass-weighted average \footnote{It can also be calculated using volume-weighted average as $\widetilde{q} = \overline{\rho q}/\overline{\rho}$. }. defined as

\begin{align}\label{eq:fht}
\fht{q}(x,t) = \frac{1}{T\Delta M}\int_{t- T/2}^{t+T/2} q(x,y,z,t')~dMdt'
\end{align}

where $dM = \rho dV$ and $M$ is $\rho V$. Thus, overbar represents the so-called Reynolds average and tilde represent a so-called Favre average. For a more complete elaboration on the algebra of these averaging procedures we refer the reader to \citet{Chassaing2010}.

\section{Snapshots of composition distribution}
\label{ap:snapshots}

\begin{figure*}
\includegraphics[width=0.45\hsize]{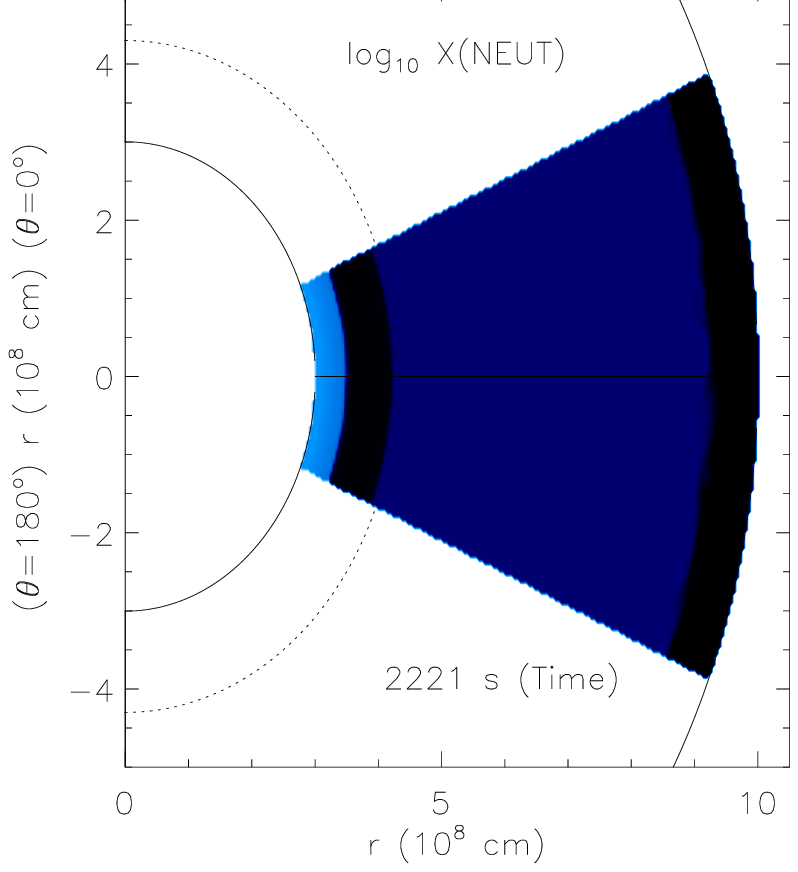}
\includegraphics[width=0.43\hsize]{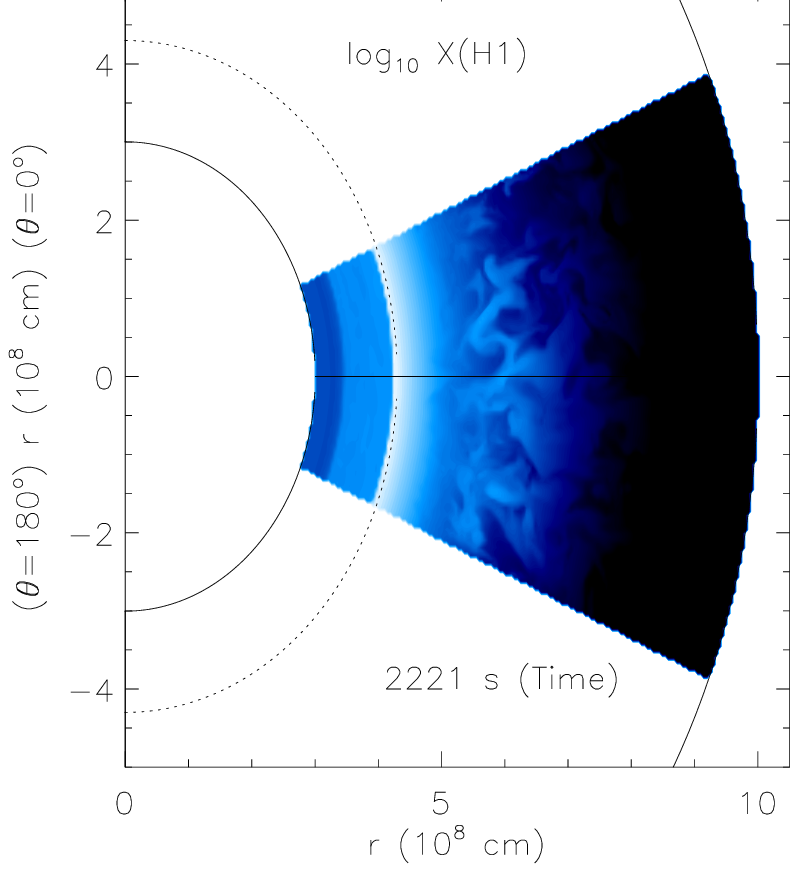} \\

\includegraphics[width=0.43\hsize]{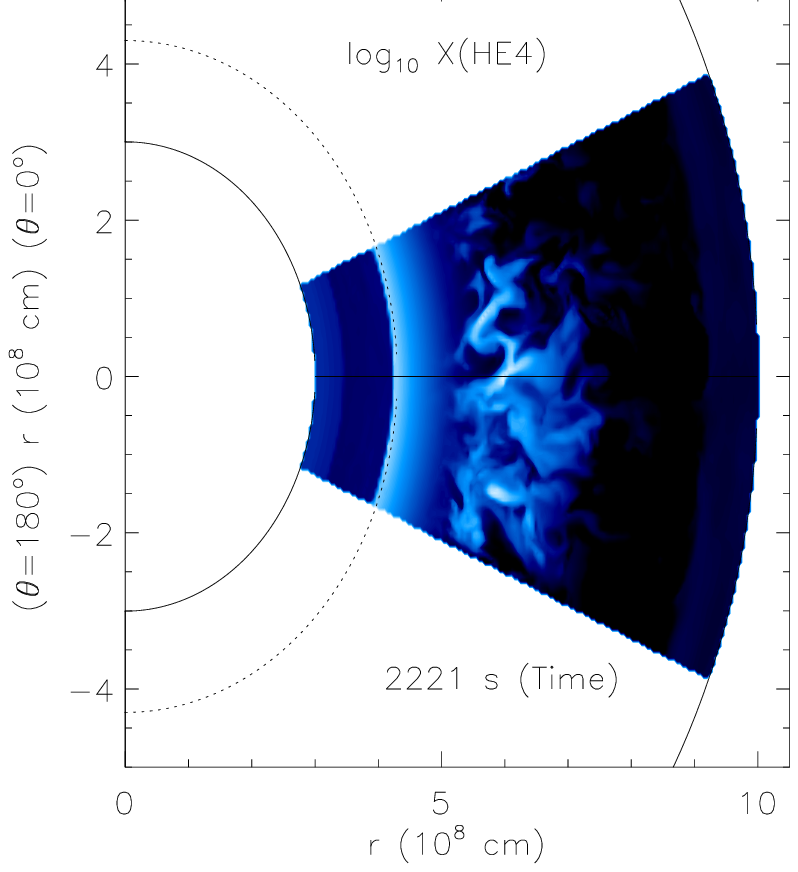}
\includegraphics[width=0.43\hsize]{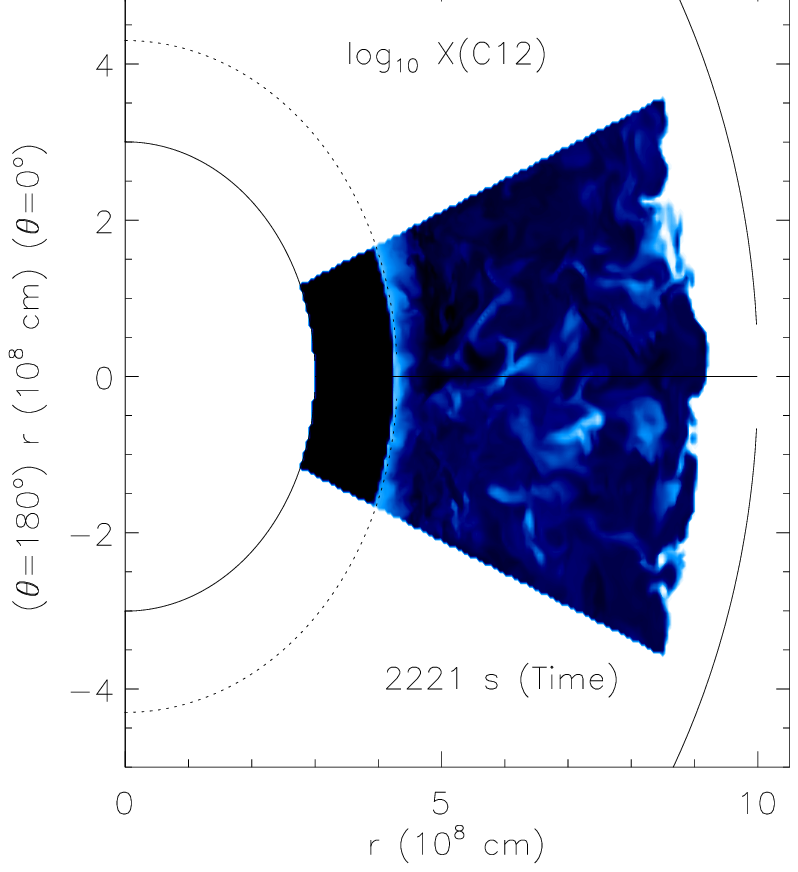} \\

\includegraphics[width=0.43\hsize]{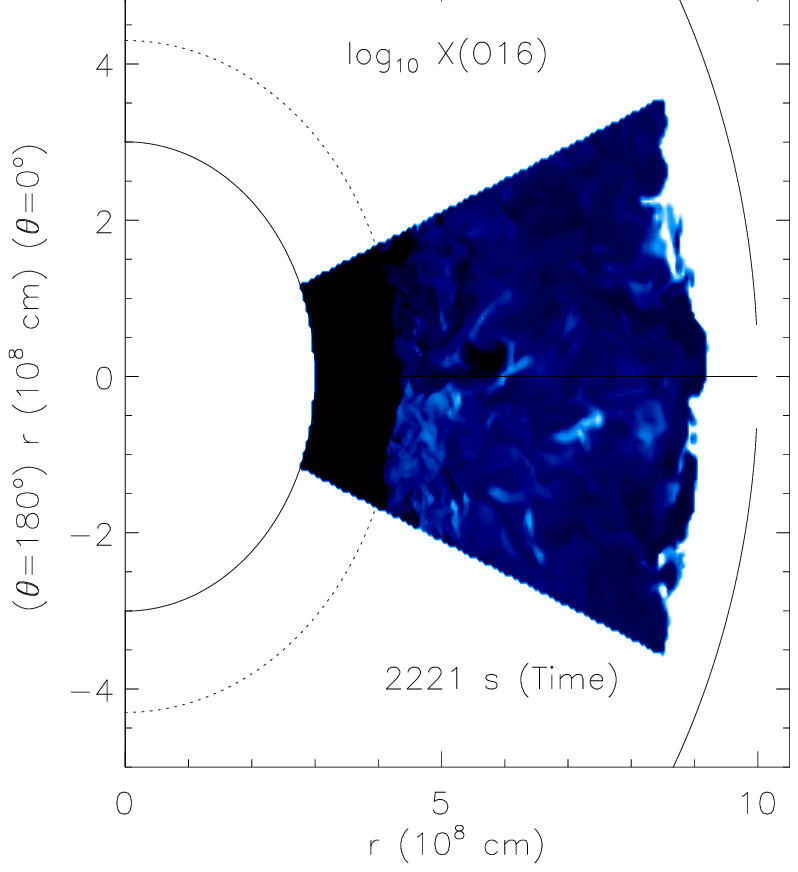}
\includegraphics[width=0.43\hsize]{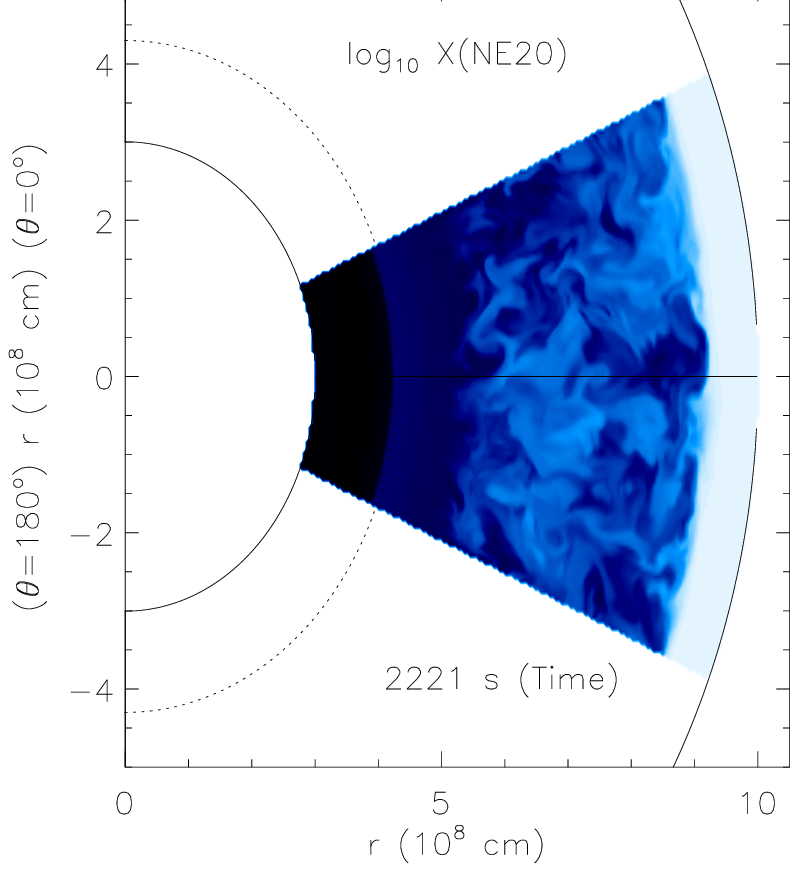}
\caption{Cuts through a 3D domain showing mass fraction of various chemical elements.
Dashed line marks location of temperature maximum. Dotted lines mark grid boundaries.}
\label{fig:2dcuts_neuth1he4c12o16ne20}
\end{figure*}

\begin{figure*}
\includegraphics[width=0.43\hsize]{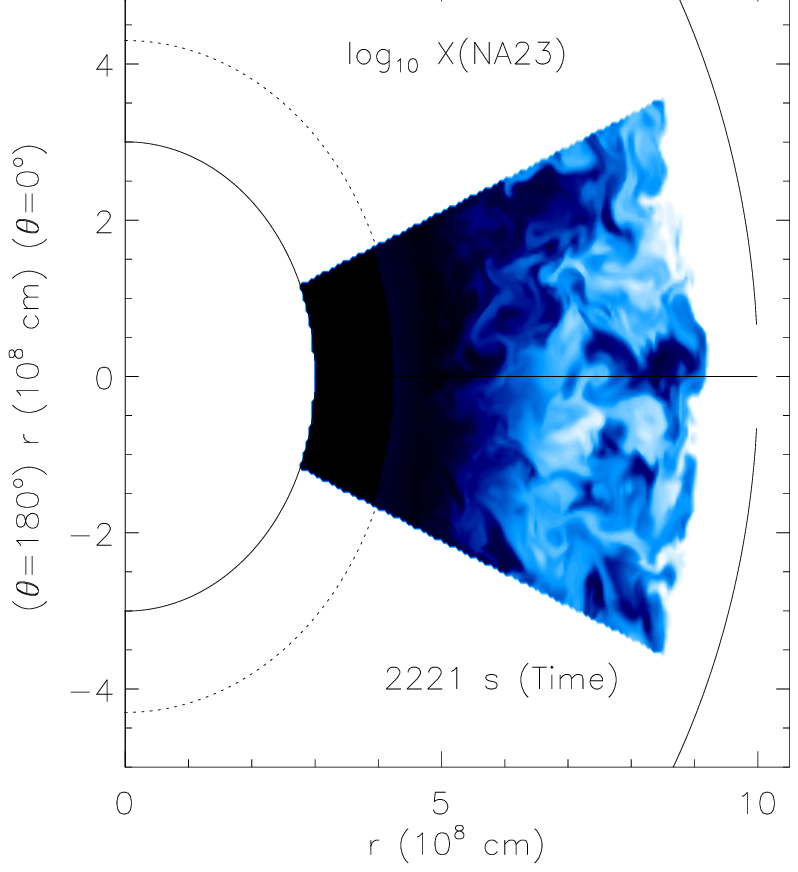}
\includegraphics[width=0.43\hsize]{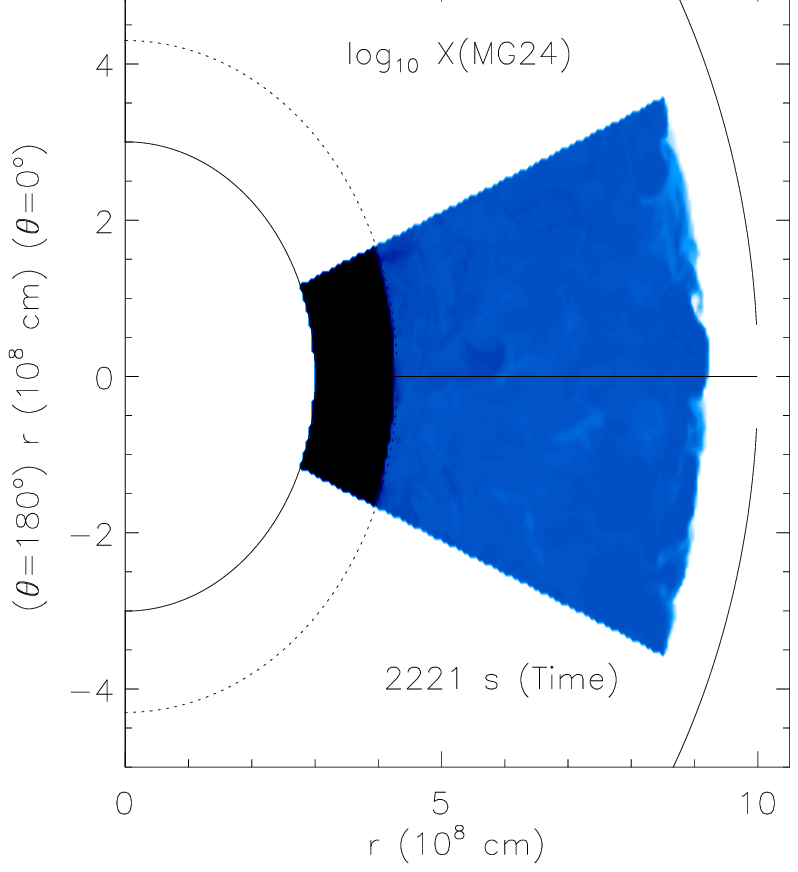} \\

\includegraphics[width=0.43\hsize]{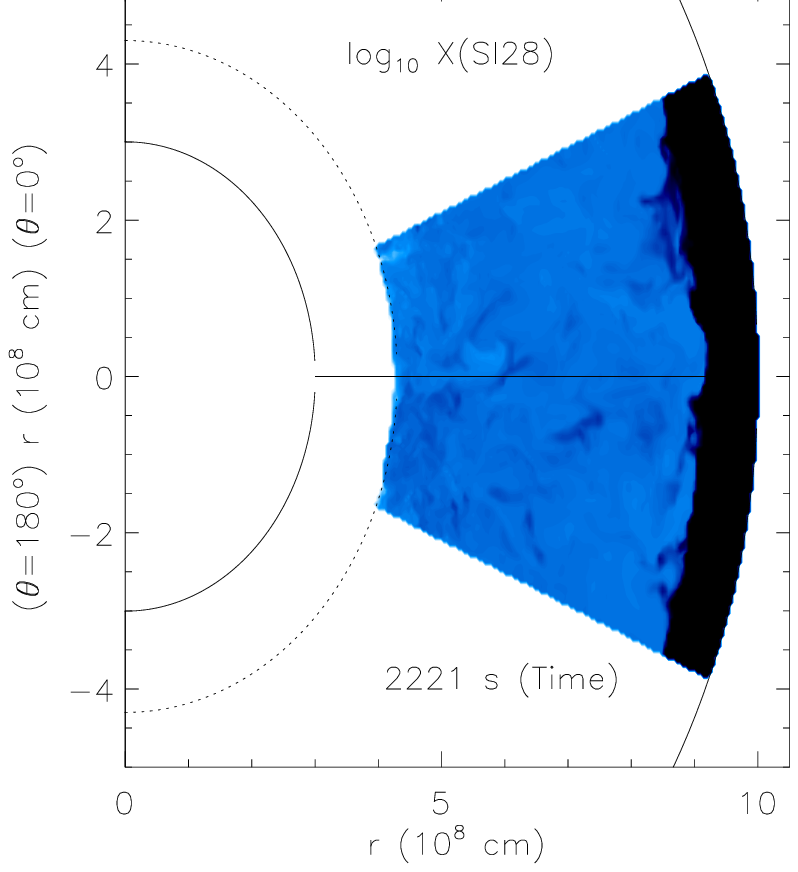}
\includegraphics[width=0.43\hsize]{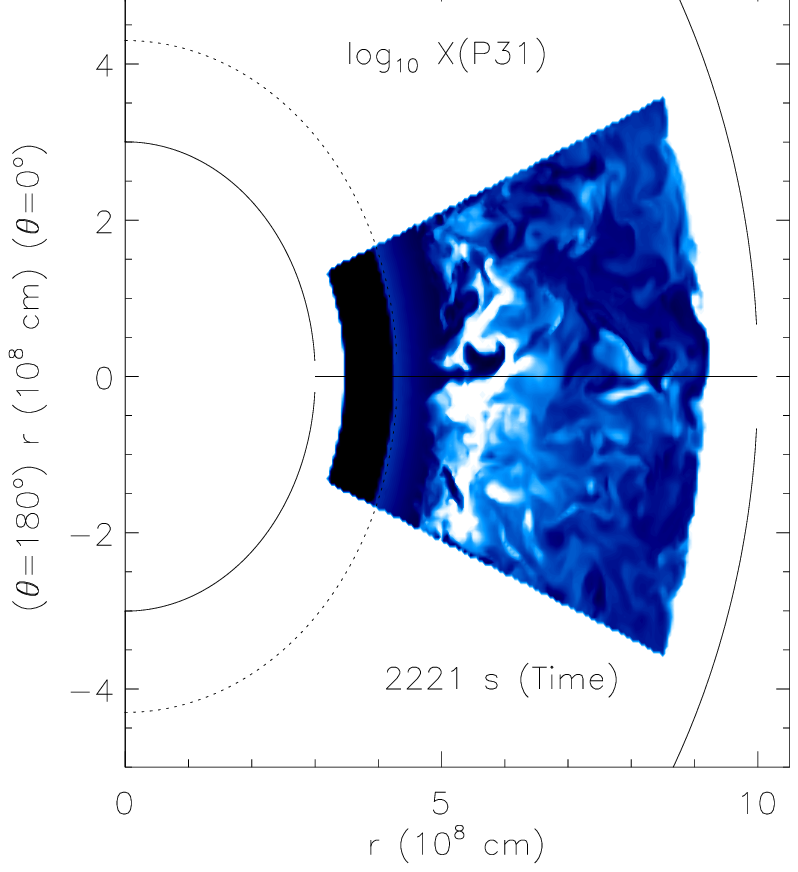} \\

\includegraphics[width=0.43\hsize]{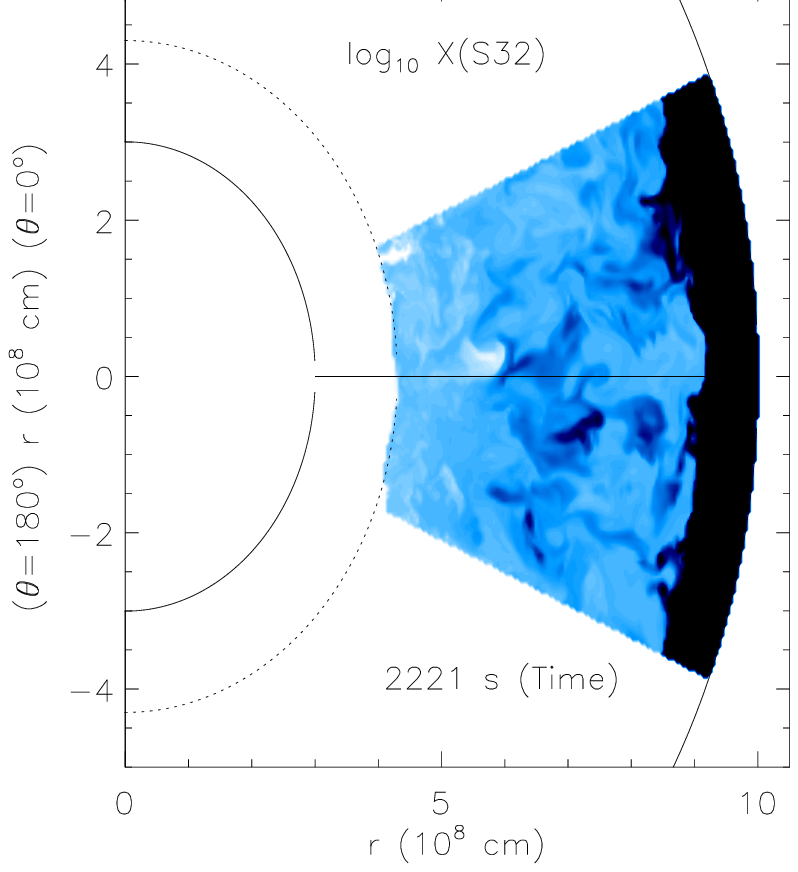}
\includegraphics[width=0.43\hsize]{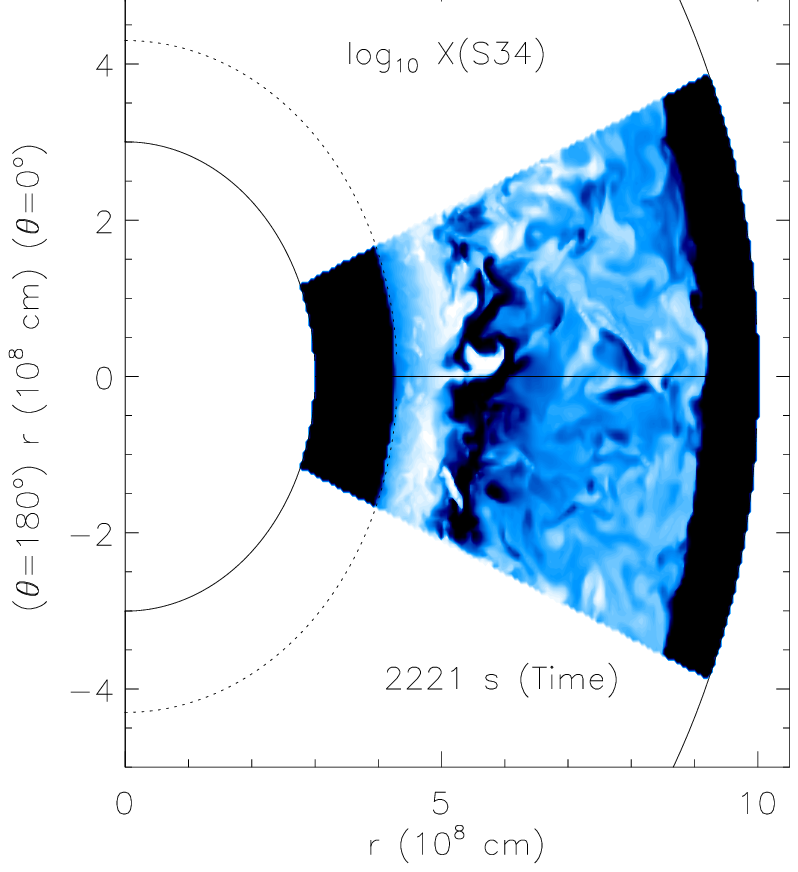}
\caption{Cuts through a 3D domain showing mass fraction of various chemical elements.
Dashed line marks location of temperature maximum. Dotted lines mark grid boundaries.}
\label{fig:2dcuts_na23mg24si28p31s32s34}
\end{figure*}

\begin{figure*}
\includegraphics[width=0.43\hsize]{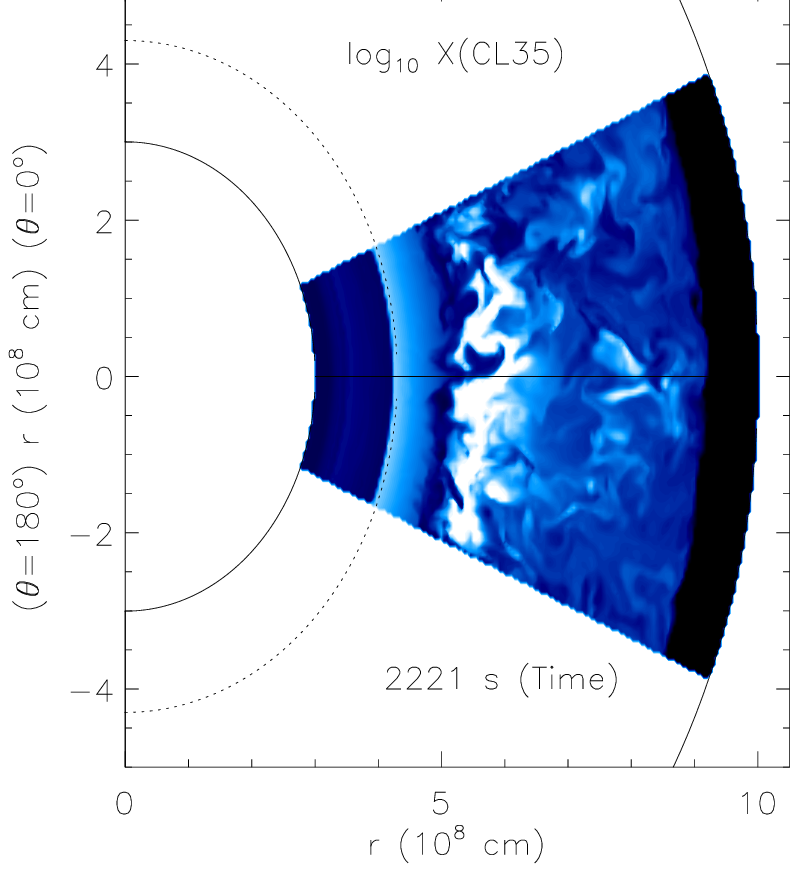}
\caption{Cuts through a 3D domain showing mass fraction of various chemical elements.
Dashed line marks location of temperature maximum. Dotted lines mark grid boundaries.}
\label{fig:2dcuts_cl35}
\end{figure*}




\bibliographystyle{mnras}
\bibliography{referenc} 

\bsp	
\label{lastpage}
\end{document}